\documentclass[fleqn,10pt]{wlscirep}
\usepackage{mathtools}
\usepackage{floatrow}
\usepackage{graphicx}
\usepackage{dcolumn}
\usepackage{bm}
\usepackage{tikz}
\usetikzlibrary{shapes, arrows, calc, positioning,matrix}
\usepackage[export]{adjustbox}
\usepackage{bbold}
\usepackage{graphicx}
\usepackage{hyperref}
\usepackage[font=small,justification=RaggedRight,labelfont=bf]{caption}
\usepackage{enumitem}
\usepackage[utf8]{inputenc}
\usepackage[T1]{fontenc}
\usepackage{soul,xcolor}
\usepackage{float}
\usepackage[label font=bf,labelformat=simple]{subfig}
\usepackage{multirow}

\makeatletter
\newcommand{\subalign}[1]{%
  \vcenter{%
    \Let@ \restore@math@cr \default@tag
    \baselineskip\fontdimen10 \scriptfont\tw@
    \advance\baselineskip\fontdimen12 \scriptfont\tw@
    \lineskip\thr@@\fontdimen8 \scriptfont\thr@@
    \lineskiplimit\lineskip
    \ialign{\hfil$\m@th\scriptstyle##$&$\m@th\scriptstyle{}##$\crcr
      #1\crcr
    }%
  }
}
\makeatother

\title{Maximum Entropy approach to multivariate time series randomization}

\author[1]{Riccardo Marcaccioli}
\author[1,2,*]{Giacomo Livan}
\affil[1]{Department of Computer Science, University College London, 66-72 Gower Street, London WC1E 6EA, UK}
\affil[2]{Systemic Risk Centre, London School of Economics and Political Sciences, Houghton Street, London WC2A 2AE, UK}
\affil[*]{g.livan@ucl.ac.uk}

\begin{abstract}
Natural and social multivariate systems are commonly studied through sets of simultaneous and time-spaced measurements of the observables that drive their dynamics, i.e., through sets of time series. Typically, this is done via hypothesis testing: the statistical properties of the empirical time series are tested against those expected under a suitable null hypothesis. This is a very challenging task in complex interacting systems, where statistical stability is often poor due to lack of stationarity and ergodicity. Here, we describe an unsupervised, data-driven framework to perform hypothesis testing in such situations. This consists of a statistical mechanical approach - analogous to the configuration model for networked systems - for ensembles of time series designed to preserve, on average, some of the statistical properties observed on an empirical set of time series. We showcase its possible applications with a case study on financial portfolio selection.
\end{abstract}
\begin{document}

\flushbottom
\maketitle
%
%
\thispagestyle{empty}

\section{Introduction}

Hypothesis testing lies at the very core of the scientific method. In its general formulation, it hinges upon contrasting the observed statistical properties of a system with those expected under a null hypothesis. In particular, hypothesis testing allows to discard potential models of a system when empirical measurement that would be exceedingly unlikely under them are made.

However, there is often no theory to guide the investigation of a system's dynamics. What is worse, in many practical situations one may be given a single - and possibly unreproducible - set of experimental data. This is indeed the case when dealing with most complex systems, whose collective dynamics often are markedly non-stationary, ranging from climate \cite{non-stationarity-climate,non-stationarity-climate-2} to brain activity \cite{non-stationarity-brain} and financial markets \cite{tsallis2003nonextensive,non-stationarity-market-cont,non-stationarity-market-livan}. This, in turn, makes hypothesis testing in complex systems a very challenging task, that potentially prevents from assessing which properties observed in a given data sample are ``untypical'', i.e, unlikely to be observed again in a sample collected at a different point in time.  

This issue is usually tackled by constructing ensembles of artificial time series sharing some characteristics with those generated by the dynamics of the system under study. This can be done either via modelling or in a purely data-driven way. In the latter case, the technique most frequently used by both researchers and practitioners is bootstrapping~\cite{bootstrap1997,bootstrap2018introduction}, which amounts to generating partially randomised versions of the available data via resampling, that can then be used as a null benchmark to perform hypothesis testing. Depending on its specificities, bootstrapping can account for autocorrelations and cross-correlations in time series sampled from multivariate systems. However, it relies on assumptions, such as sample independence and some form of stationarity~\cite{bootstrap_lim}, which limit its power when dealing with complex systems. 

As far as model-driven approaches are concerned, the literature is extremely vast~\cite{uber_omnicomprensive_ref}. Broadly speaking, modelling approaches rely on a priori structural assumptions for the system's dynamics, and on identifying the parameter values that best explain the available set of observations within a certain class of models (e.g., via Maximum Likelihood~\cite{uber_omnicomprensive_ref}). A widely used class for multivariate time series is that of autoregressive models, such as VAR~\cite{VAR}, ARMA~\cite{ARMA}, and GARCH~\cite{garch}, which indeed were originally introduced, among other reasons, to perform hypothesis testing~\cite{ARMA}. In such models, the future values of each time series are given by a linear combination of past values of one or more time series, each characterised by their own idiosyncratic noise to capture the fluctuations of individual variables. Such a structure is most often dictated by its simplicity rather than by first principles. As a consequence, once calibrated, autoregressive models produce rather constrained ensembles of time series that do not allow to explore scenarios that differ substantially from those observed empirically.

Another modelling philosophy places more emphasis on capturing the structural collective properties of multivariate systems rather than their dynamical ones. Random Matrix models are a prime example in this direction, which usually rely on ansatzes on the correlation structure of the system under study trying to strike a balance between the resulting models' analytical tractability and their adherence to empirical observations. One of the first - and still most widely used - Random Matrix models is the Wishart Ensemble~\cite{wishart1948proofs,livan2018introduction}, which in its  simplest form leads to the much celebrated Mar\v cenko-Pastur distribution~\cite{marvcenko1967distribution} for uncorrelated systems, up to rather recent developments to tackle non-stationarities in financial data~\cite{schmitt2013non}.

Here we propose a Maximum Entropy approach - inspired by Statistical Mechanics - to perform hypothesis testing on sets of time series. Starting from the Maximum Entropy principle, we will introduce a (gran)canonical ensemble of correlated time series subject to constraints based on the properties of an empirically observed set of measurements. This, in turn, will result in a multivariate probability distribution which allows to unbiasedly sample values centred on such measurements, which represents the main contribution of this paper. The theory we propose in the following shares some similarities with the canonical ensemble of complex networks~\cite{canonical_ens_net_newman,can_ens_continuous,can_ens_nature_rev,masuda2018configuration}, and, as we will show, inherits its powerful calibration method based on Likelihood maximization~\cite{can_ens_squartini}. 

The paper is organized as follows. In the next Section, we outline the general formalism of our approach. Then, as a formative example, we show how the methodology introduced can be used to reconstruct an unknown probability density function from repeated measurements over time. After that, we proceed to study the most general case of multivariate time series, showing how our approach recovers collective statistical properties of interacting systems without directly accounting for such interactions in the set of constraints imposed on the ensemble. Before concluding with some final remarks, we briefly mention an interesting analogy between our approach and Jaynes' Maximum Caliber principle \cite{jaynes1980minimum}. 

\section*{General Framework Description}

\begin{figure}
\centering
  \includegraphics[width=0.9\textwidth]{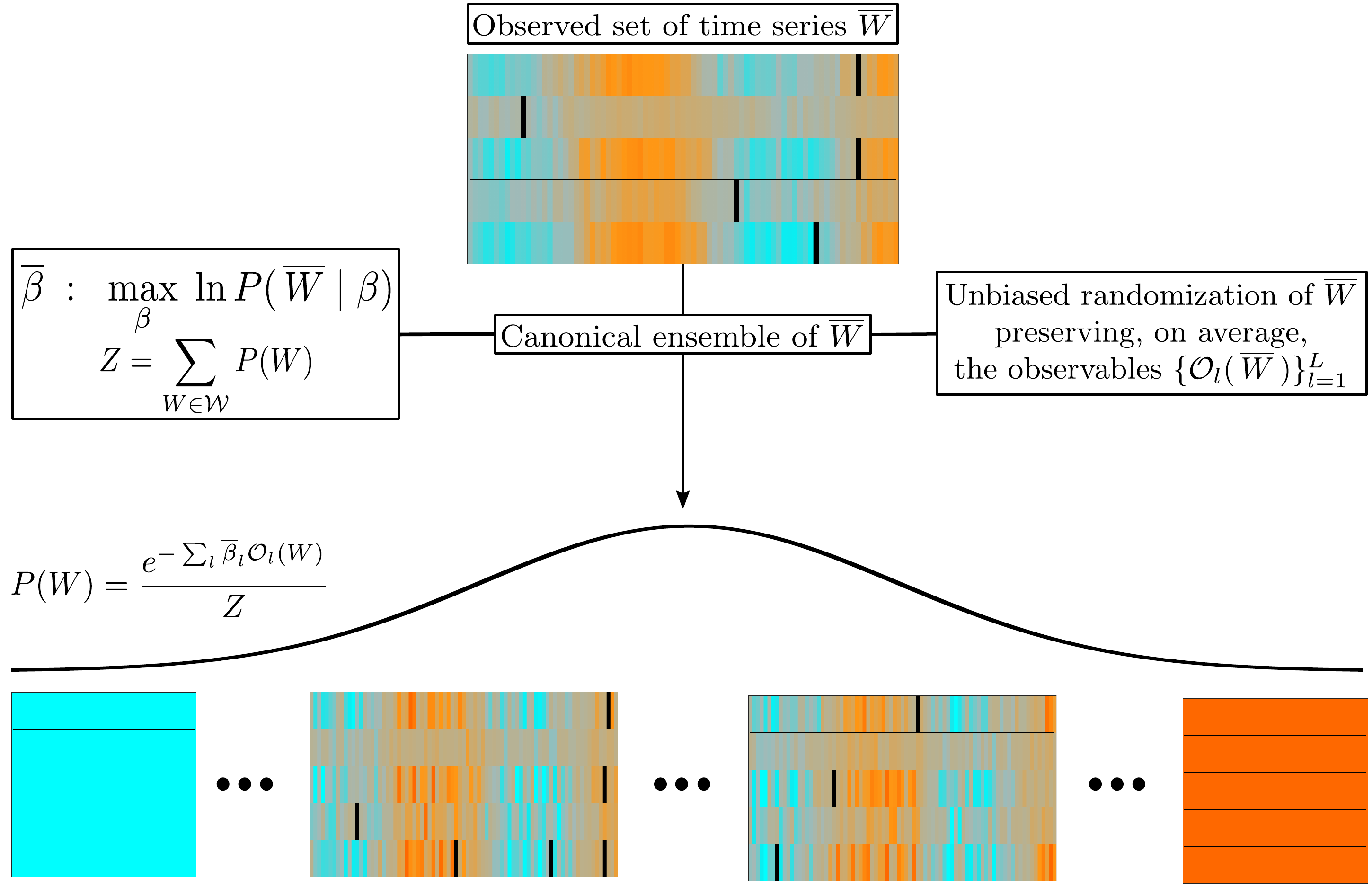}
  \caption{\textbf{Schematic representation of the model.} Starting from an empirical set of time series $\overline{W}$, we construct its unbiased randomization by finding the probability measure $P(W)$ on the phase space $\mathcal{W}$ which maximises Gibbs' entropy while preserving the constraints $\{\mathcal{O}_l(\overline{W}) \}_{l=1}^L$ as ensemble averages. The probability distribution $P(W)$ depends on $L$ parameters that can be found by maximising the likelihood of drawing $\overline{W}$ from the ensemble. In the Figure, orange, turquoise and black are used to indicate  positive, negative or empty values of the entries $W_{i t}$, respectively, while brighter shades of each color are used to display higher absolute values. As it can be seen, the distribution $P(W)$ assigns higher probabilities to those sets of time series that are more consistent with the constraints and therefore more similar to $\overline{W}$. See \cite{can_ens_nature_rev} for a similar chart in the case of the canonical ensemble of complex networks.}
  \label{fig: model representation}
\end{figure}


Let $\mathcal{W}$ be the set of all real-valued sets of $N$ time series of length $T$ (i.e., the set of real-valued matrices of size $N \times T$), and let $\overline{W} \in \mathcal{W}$ be the empirical set of data we want to perform hypothesis testing on (i.e., $\overline{W}_{i t}$ stores the value of the $i$-th variable in the system sampled at time $t$, so that $\overline{W}_{i t}$ for $t = 1, \ldots, T$ represents the sampled time series of variable $i$). Our aim is to define a probability density function $P(W)$ on $\mathcal{W}$ such that the expectation values $ \langle \mathcal{O}_\ell(W) \rangle $ of a set of observables ($\ell = 1, \ldots, L$) coincide with the value $\overline{O}_\ell$ of the corresponding quantity as empirically measured from $\overline{W}$. 

Following Boltzmann and Gibbs, we can achieve the above by invoking the Maximum Entropy principle, i.e., by identifying $P(W)$ as the distribution that maximises the entropy functional $S(W) = - \sum_{W \in \mathcal{W}} P(W) \, \ln  P(W)$, while satisfying the $L$ constraints $\langle \mathcal{O}_\ell (W) \rangle = \sum_W \mathcal{O}_\ell(W) P(W) = \overline{O}_\ell$ and the normalisation condition $\sum_W P(W) = 1$. As is well known \cite{jaynes1957information,jaynes1957informationII}, this reads
\begin{equation}
    P(W) = \frac{e^{- H(W)}}{Z} \ ,
\end{equation}
where $H(W) = \sum_\ell \beta_\ell \ \mathcal{O}_\ell(W)$ is the Hamiltonian of the system, $\beta_\ell$ ($\ell = 1, \ldots, L$) are Lagrange multipliers introduced to enforce the constraints, and $Z =\sum_W e^{- H(W)}$ is the partition function of the ensemble, which verifies $ \langle \mathcal{O}_\ell(W) \rangle = \partial \ln Z / \partial \beta_\ell \; , \forall \, \ell $.

Figure~\ref{fig: model representation} provides a sketch representation of the ensemble theory just introduced. The rationale of enforcing the aforementioned constraints is that of finding a distribution $P(W)$ that assigns low probability to regions of the phase space $\mathcal{W}$ where the observables associated to the Lagrange multipliers $\beta_\ell$ take values that are exceedingly different from those measured in the empirical set $\overline{W}$, and high probability to regions where some degree of similarity with $\overline{W}$ is retained (it should be noted that in some cases this does not necessarily leads to the distribution $P(W)$ being peaked around the values $\overline{O}_\ell$). This, in turn, allows to test whether other properties (not encoded in any of the constraints) of $\overline{W}$ are statistically significant by measuring how often they appear in instances drawn from the ensemble. The existence and uniqueness of the Lagrange multipliers ensuring the ensemble's ability to preserve the constraints $\overline{O}_\ell$ is a well known result, and they are equivalent to those that would be obtained from the maximization of the Likelihood of drawing the empirical matrix $\overline{W}$ from the ensemble~\cite{lagr_mult_coincides_maximum_like}.

In the two following Sections, we shall illustrate our general framework on two examples -- one devoted to the single time series case, one to the multivariate case. In both examples, we shall make a selection of possible constraints that can be analytically captured by the approach, i.e., constraints for which the resulting ensemble's partition function can be computed in closed form. In particular, since we will apply our approach to financial data in a later section, we will choose constraints that have a clear interpretation in the analysis of financial time series. However, it should be kept in mind that such constraints are by no means to be interpreted as general prescriptions, and all results presented in the following could be reobtained -- depending on the applications of interest -- with any other set of constraints allowing for analytical solutions.

\section*{Single time series}

As a warm up example to showcase our approach, we shall consider a simple case of a univariate and stationary system with no correlations over time. This amounts to a time series made of independent and identically distributed random draws from a probability density function, which we aim at reconstructing. In the following section we will then proceed to consider multivariate and correlated cases.

Let us then consider an $1 \times T$ empirical data matrix $\overline{W}$ coming from $T$ repeated sampling of an observable of the system under consideration. If the processes is stationary and time-independent, this is equivalent to sampling $T$ times a random variable from its given, unknown, distribution and therefore the task of the model can be translated into reconstructing the unknown distribution given the data. Let us consider a vector $\xi \in [0,1]^d$ and the associated empirical $\xi$-quantiles $\overline{q}_{\xi}$ calculated on $\overline{W}$. In order to fully capture the information present in the data $\overline{W}$, we are going to constrain our ensemble to preserve, as averages, one or more quantities derived from $\overline{q}_{\xi}$. Possible choices may be:
\begin{itemize}
    \item The number of data points falling within each pair of empirically observed adjacent quantiles: \\ $\overline{N}_{\xi_i} = \sum_t \Theta(\overline{W}_t-\overline{q}_{\xi_{i-1}}) \; \Theta(-\overline{W}_t+\overline{q}_{\xi_{i}})$ 
    \item The cumulative values of the data points falling within each pair of adjacent quantiles: \\ $\overline{M}_{\xi_i} = \sum_t \overline{W}_t \; \Theta(\overline{W}_t-\overline{q}_{\xi_{i-1}}) \; \Theta(-\overline{W}_t+\overline{q}_{\xi_{i}})$
    \item The cumulative squared values of the data points falling within each pair of adjacent quantiles: \\ $\overline{M}_{\xi_i}^2 = \sum_t \overline{W}_t^2 \; \Theta(\overline{W}_t-\overline{q}_{\xi_{i-1}}) \; \Theta(-\overline{W}_t+\overline{q}_{\xi_{i}})$
\end{itemize}
In each of the above constraints we assumed $i=2,\ldots,d$, and we have used $\Theta(\cdot)$ to indicate Heaviside's step function (i.e., $\Theta(x) = 1$ for $x > 0$, and $\Theta(x) = 0$ otherwise). In general we are not required to use the same $\alpha$ for all constraints, for example we can freely choose to impose on the ensemble the ability to preserve $\overline{N}_{\xi_i} \forall i \in [1,d]$ together with the total cumulative values $\overline{M} = \sum_i \overline{M}_{\xi_i}$, and total cumulative squared values $\overline{M}^2 = \sum_i \overline{M}_{\xi_i}^2$, as well as each $\overline{M}_{\xi_i}$ and $\overline{M}_{\xi_i}^2$ separately. Note that the first constraint in the above list effectively amounts to constraining the ensemble's quantiles.

As we will discuss more extensively later on, the above set of constraints is discretionary. A possible strategy to get rid of such discretionality, would be to partition the data based on a binning procedure aimed at compromising between resolution and relevance, according to its definition introduced in Ref.~\cite{cubero2019statistical}.

A defined set of constraints will lead to a different Hamiltonian, to a different number of Lagrange multipliers and therefore to a different statistical model. If we choose, for example, to adopt all the constraints specified above, the Hamiltonian $H$ of the ensemble will depend on a total of $3 (d-1)$ parameters:
\begin{equation}
    H(W) =  \sum_{i=1}^{d} \sum_{t=1}^{T} \; \left[\, a_i + W_t \alpha_i + W^2_t \beta_i \, \right] \Theta(\overline{W}_t-\overline{q}_{\xi_{i-1}}) \; \Theta(-\overline{W}_t+\overline{q}_{\xi_{i}}) \; .
\end{equation}
The freedom to choose the amount of constraints of course comes with a cost. First of all, it must be noted that the Likelihood of the data matrix $\overline{W}$ will be in general a non linear function of the Lagrange multipliers and therefore of the constraints. These latter can vary both in magnitude (by choosing different values for the entries of $\alpha$) and in size (by choosing a different $d$). In general, tackling the issue of finding the optimal positions for the constraints, given their number $d$, can become highly not trivial and goes out of the scope of the present work. However, loosely speaking, the Likelihood of finding $\overline{W}$ after a random draw from the defined ensemble is an increasing function of the number constraints, coherently with the idea that a larger number of parameter leads to better statistics on the data used to train the model. As a result, in order to avoid overfitting, given a set of constraints, we can compare different values of $d$ by using standard model selection techniques such as the Bayesian \cite{kass1995reference} or Akaike information criteria \cite{akaike1998information}. 

In the following we are going to show how to apply the methodology just outlined to a synthetic dataset. For this example, let us assume that the data generating process follows a balanced mixture of a truncated standard Normal distribution and a truncated Student's t-distribution with $\nu =5$ degree of freedom. The two models we are going to use to build the respective ensembles are specified by the following Hamiltonians:
\begin{equation}
\label{eq:Hamiltonians}
\begin{aligned}
    & H_1 = \sum_i \left[ \alpha_i N_{\xi_i} + \beta_i M_{\xi_i} \right] \\
    & H_2 =  \sum_i \left[ \alpha_i N_{\xi_i} + \beta M_{\xi_i} + \gamma M^2_{\xi_i} \right]
\end{aligned}
\end{equation}
The model resulting from $H_1$ will have a total of $2(d-1)$ parameter and will preserve the average number of data point contained within each pair of adjacent quantiles together with their cumulative values, while the model coming from $H_2$ will be characterised by $d+1$ parameters and will preserve the average number of data point contained within each pair of adjacent quantiles together with the overall mean and variance calculated across all data points. In order to find the Lagrange multipliers able to preserve the chosen constraints, we first need to the find the partitions functions of the two ensembles $Z_{1,2} = \sum_{W} e^{- H_{1,2}(W)}$. In order to do that, we first need to specify the sum over the phase space:
\begin{equation}
    \sum_{W \in \mathcal{W}}  \equiv \prod_{t=1}^{T} \int_{\overline{q}_{\xi_1}}^{\overline{q}_{\xi_d}} d w_{t} \ .
\end{equation}
The above expression leads to the following partition functions (in Appendix \ref{app:derivation} we present a detailed derivation of the partition function shown in the following Section; the following result can be obtained with similar steps):
\begin{align}
    & Z_1 = \prod_{t=1}^{T} \sum_{i=1}^{d-1} e^{-\alpha_i} \frac{e^{-\beta_i \overline{q}_{\xi_i}}-e^{-\beta_i \overline{q}_{\xi_{i+1}}}}{\beta_i} \\
    & Z_2 = \prod_{t=1}^{T} \sum_{i=1}^{d-1} \sqrt{\frac{\pi}{4 \gamma}} \; e^{\frac{\beta^2}{4 \gamma}-\alpha_i} \left( - \text{erf}\left[ \frac{\beta + 2 \gamma \overline{q}_{\xi_i}}{2 \sqrt{\gamma}} \right] + \text{erf}\left[ \frac{\beta + 2 \gamma \overline{q}_{\xi_{i+1}}}{2 \sqrt{\gamma}} \right] \right)
\end{align}
where with $\text{erf}$ we indicate the Gaussian error function $\text{erf}(z) = \frac{2}{\pi} \int_{0}^{z} e^{-t^2} dt$. 

\begin{figure}
  \centering
  \includegraphics[width=1\linewidth]{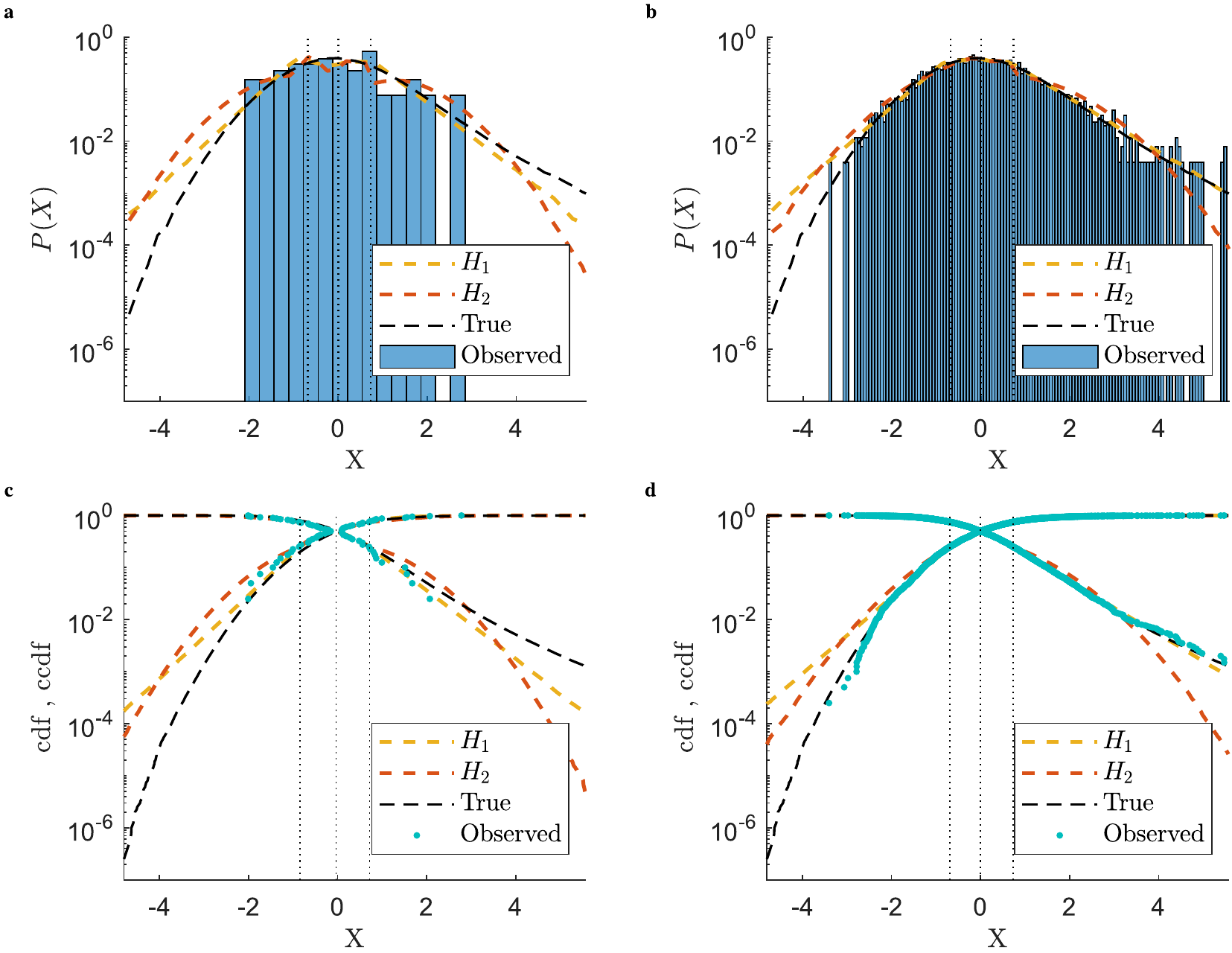}
  \caption{Comparisons between empirical PDFs (shown as histograms), CDFs and survival functions (CCDFs) and their empirical counterparts reconstructed with our ensemble approach from the Hamiltonians in Eq.~\eqref{eq:Hamiltonians}, shown in yellow ($H_1$) and red ($H_2$), respectively. \textbf{a)} Results on the PDF obtained by calibrating the models on 40 data points. \textbf{b)} Results on the PDF obtained by calibrating the models on 4000 data points. \textbf{c}) Results on the CDF (and associated survival function) with models calibrated on 40 data points. \textbf{d}) Results on the CDF (and associated survival function) with models calibrated on 4000 data points. In all plots the dashed black lines marked as ``True'' correspond to the analytical PDF, CDF and survival function (depending on the panel) of the synthetic data generating process (given by a mixture of a Gaussian and a Student-$t$, see main text). Vertical dashed lines correspond to the 0.25, 0.5, and 0.75 quantiles employed to calibrate the models. The results from the ensemble have been obtained by pooling together $10^6$ time series independently generated from the ensemble.}
\label{fig: pdf reconstruction}
\end{figure}

In Figure~\ref{fig: pdf reconstruction} we show how the models resulting from the partition functions $Z_1$ and $Z_2$ are able to reconstruct the underlying true distribution starting from different amount of information (i.e different sample sizes) and quantiles vector set to $\overline{q} = [-\infty,\overline{q}_{0.25},\overline{q}_{0.5},\overline{q}_{0.75},\infty]$ (the steps to obtain the ensemble's probability density function from its partition function are outlined in Appendix \ref{app:derivation} for the case discussed in the next Section). First of all we note that, as expected, estimating the unknown distribution from more data gives estimates that are closer to the real underlying distribution. Moreover, looking at Figure~\ref{fig: pdf reconstruction}, we can qualitatively see that the model described by $Z_1$ does a better job than $Z_2$ at inferring the unknown pdf. We can verify both statements more quantitatively by calculating the the Kullback–Leibler divergence of the estimated distributions from the true one: for the case with 40 data points we observe $D_{KL}(P_{Z_1} | P_{T}) = 0.10$ and $D_{KL}(P_{Z_2} | P_{T}) = 0.19$ while for the case with 4000 samples we have $D_{KL}(P_{Z_1} | P_{T}) = 0.01$ and $D_{KL}(P_{Z_2} | P_{T}) = 0.08$. Of course, we cannot conclude yet that $Z_1$ gives overall a better model for our reconstruction task than $Z_2$ since they are described by a different number of parameters. As mentioned above, in order to complete our model comparison exercise, we need to rely on a test able to assess the relative quality of the models for a given set of data. We choose the Akaike information criterion which uses as its score function the following $\text{AIC} = 2 k - 2 \log \hat{L}$, where $k$ is the number of estimated parameters and $\hat{L}$ is the maximum value of the likelihood function for the model. We end up with $\text{AIC}_{Z_1} = 130$ and $\text{AIC}_{Z_2} = 950$ for the 40 data points case and $\text{AIC}_{Z_1} = 1,15 \times 10^4$ and $\text{AIC}_{Z_2} = 2.11 \times 10^{5}$ when 4000 data points are available. Of course it is worth repeating that all the steps mentioned above are performed given a fixed vector $\overline{q}$ common to the two models.

\floatsetup[figure]{style=plain,subcapbesideposition=top}

\section*{Multiple time series}

In this Section we proceed to present the application of the approach introduced above to the multivariate case.

Let us consider an $N \times T$ empirical data matrix $\overline{W}$ whose rows have been rescaled to have zero mean, so that $\overline{W}_{it} > 0$ ($\overline{W}_{it} < 0$) will indicate that the time $t$ value of the $i$-th variable is higher (lower) than its empirical mean. Also, without loss of generality, let us assume that $\overline{W}_{it} \in \mathbb{R}_{\ne 0}$, and that $\overline{W}_{it} = 0$ indicates missing data. For later convenience, let us define  $A^{\pm} = \Theta(\pm W)$ and $w^{\pm} = \pm W \Theta (\pm W)$ (we shall denote the corresponding quantities measured on the empirical set as $\overline{A}^{\pm}$ and $\overline{w}^{\pm}$), and let us constrain the ensemble to preserve the values of the following observables:
\begin{itemize}
    \item The number of positive (above-average) and negative (below-average) values $\overline{N}_i^\pm = \sum_t \overline{A}^{\pm}_{it}$, and the number of missing values $\overline{N}_i^0 = T - \overline{N}_i^+ - \overline{N}_i^-$ recorded for each time series ($i = 1, \ldots, N$).
    \item The cumulative positive and negative values recorded for each time series: $\overline{S}_i^\pm = \sum_t \overline{w}_{it}^\pm$ ($i = 1, \ldots, N$).
    \item The number of positive, negative, and missing values recorded at each sampling time: $\overline{M}_t^\pm = \sum_i \overline{A}^\pm_{it}$, $\overline{M}_t^0 = N - \overline{M}_t^+ - \overline{M}_t^-$ ($t = 1, \ldots, T$).
    \item The cumulative positive and negative value recorded at each sampling time: $\overline{R}_t^\pm = \sum_i \overline{w}^\pm_{it}$ ($t = 1, \ldots, T$).
\end{itemize}
Note that the second constraint in the above list indirectly constrains the mean of each time series. As mentioned in the General Framework section, we selected the above constraints inspired by potential financial applications (and indeed we will assess the ensemble's ability to capture time series behaviour on financial data). In such context, the above four constraints respectively correspond to: the number of positive and negative returns of a given financial stock, the total positive and negative return of a stock, the number of stocks with a positive or negative return on a given trading day, the total positive and negative return across all stocks on a given trading day. Such constraints amount to some of the most fundamental ``observables'' associated with financial returns. As we will detail in the following, forcing the ensemble to preserve them on average also amounts to effectively preserving other quantities that are of paramount importance in financial analysis, such as, e.g., the skewness and kurtosis of return distributions, and some of the correlation properties of a set of financial stocks (which are central to financial portfolio analysis and selection.)

The above list amounts to $8(N + T)$ constraints, and the Hamiltonian $H$ depends on the very same number of parameters:
\begin{equation}\label{eq: hamiltonian}
    H(W) = \sum_{i=1}^N \sum_{t=1}^T \bigg [ \left(\alpha^N_i+\alpha^T_t \right) A^+_{it} + \left( \beta^{N}_{i} + \beta^T_t \right) A^-_{it} + \left( \gamma_i^N + \gamma_t^T \right) w^+_{it} + \left( \sigma_i^{N} + \sigma_t^T \right) w^-_{it} \bigg ] \ ,
\end{equation}
where we have introduced the Lagrange multipliers associated to all constraints. This choice for the Hamiltonian naturally generalizes the framework introduced in \cite{almog2014binary}. 

Let us remark that none of the above constraints explicitly accounts for either cross-correlations between variables or for correlations in time. Accounting for these would amount to constraining products of the type $\sum_{t=1}^T w_{it} w_{jt}$ (in the case of cross-correlations) and $\sum_{i=1}^N w_{it} w_{it^\prime}$ (in the case of temporal correlations), which introduce a direct coupling between the entries of $W$, resulting in a considerable loss in terms of analytical tractability of the approach. However, as we shall see in a moment, the combination of the above constraints is enough to indirectly capture some of the correlation properties in the data of interest.

In order to calculate the partition function $Z = \sum_{W} e^{- H(W)}$, we first need to properly specify the sum over the phase space. Given the matrix representation we have chosen for the system, and the fact that $w_{it}^\pm = w_{it} A^\pm_{it}$, this reads:
\begin{equation} \label{eq: sum on config main}
    \sum_{W \in \mathcal{W}}  \equiv  
    \prod_{i=1}^{N} \prod_{t=1}^{T}  \int_{-\infty}^{+\infty} d w_{i t} = \prod_{i=1}^{N} \prod_{t=1}^{T} \sum_{\subalign{
       &(0,1)\\
       (A_{i t}^+,A_{i t}^-)=&(1,0)\\
       &(0,0)
    }} \; \int_{0}^{+\infty} d w^+_{i t} \; \int_{0}^{+\infty} d w^-_{i t}  \ ,
\end{equation}
where the sum specifies whether the entry $A_{it}$ stores a positive, negative or missing value, respectively. This signifies that negative and positive events (this in general holds for any discretization of the distribution of the entries of $W$), i.e., values above and below the empirical mean of each variable, obviously cannot coexist in an entry $W_{i t}$, which, once occupied, cannot hold any other event. In this respect, we can anticipate that negative and positive events will effectively be treated as different fermionic species populating the energy levels of a physical system. Following this line of reasoning, the role of $w_{i t}^\pm$ is that of general coordinates for each of the two fermionic species. In principle, the integrals in Eq.~\eqref{eq: sum on config main} could have as upper limits some quantities $U^\pm_{it}$ to incorporate any possible prior knowledge on the bounds of the variables of interest.

The above expression leads to the following partition function (fully derived in Appendix \ref{app:derivation}):
\begin{equation}\label{eq: partition function full}
    Z  = \sum_{W \in \mathcal{W}} e^{- H(W)} = \prod_{i=1}^{N} \prod_{t=1}^{T} Z_{it}   = \prod_{i=1}^{N} \prod_{t=1}^{T}  \left[ 1 + \frac{e^{-(\alpha^{N}_i+\alpha^{T}_t)}}{\gamma_i^{N}+\gamma_t^{T}} + \frac{e^{-(\beta^{N}_i+\beta^{T}_t)}}{\sigma_i^{N}+\sigma_t^{T}} \right] = \prod_{i=1}^{N} \prod_{t=1}^{T} \left ( 1 + e^{\frac{\mu_{it}^1 - \epsilon_{it}}{T_{it}}} + e^{\frac{\mu_{it}^2 - \epsilon_{it}}{T_{it}}} \right ) \ ,
\end{equation}
where the quantities $\mu_{it}^{1,2}$, $\epsilon_{it}$, and $T_{it}$ are functions of the Lagrange multipliers (specified in Appendix \ref{app:derivation}). 

Some considerations about Eq.~\eqref{eq: partition function full} are now in order. First of all, the partition function factorises into the product of independent factors $Z_{it}$, and therefore into a collection of $N \times T$ \emph{statistically independent} sub-systems. However, it is crucial to notice that their parameters (i.e., the Lagrange multipliers) are coupled through the system of equations specifying the constraints ($ \langle \mathcal{O}_\ell(W) \rangle = \partial \ln Z / \partial \beta_\ell \; , \forall \, \ell $). As we shall demonstrate later, this ensures that part of the original system's correlation structure is retained within the ensemble. Moreover, with the above positions, the aforementioned physical analogy becomes clear: the system described by Eq.~\eqref{eq: partition function full} can be interpreted as a system of $N \times T$ orbitals with energies $\epsilon_{it}$ and local temperatures $T_{it}$ that can be populated by fermions belonging to two different species characterised by local chemical potentials $\mu_{it}^1$ and $\mu_{it}^2$, respectively.

From the partition function in Eq.~\eqref{eq: partition function full} we can finally calculate the probability distribution $P(W)$:
\begin{equation}\label{eq: final prob}
 P(W) = \prod_{i=1}^{N} \prod_{t=1}^{T} \left[ P_{it}^+ \right]^{A^+_{it}} \left[ P_{it}^- \right]^{A^-_{it}} \left[ 1 - P_{it}^+ -P_{it}^- \right]^{1-A^+_{it}-A^-_{it}} \left[ Q_{it}^+(w^+_{it}) \right]^{A^+_{it}} \; \left[ Q_{it}^-(w^-_{it}) \right]^{A^-_{it}} \; ,
\end{equation}
where $P_{it}^\pm$ and $Q_{it}^\pm(w^\pm_{it})$ are functions of the Lagrange multipliers (specified in Appendix \ref{app:derivation}) and correspond, respectively, to the probability of drawing a positive (negative) value for the $i$-th variable at time $t$ and to its probability distribution.

\floatsetup[figure]{style=plain,subcapbesideposition=top}
\begin{figure*}
  \centering
\includegraphics[width=1\linewidth]{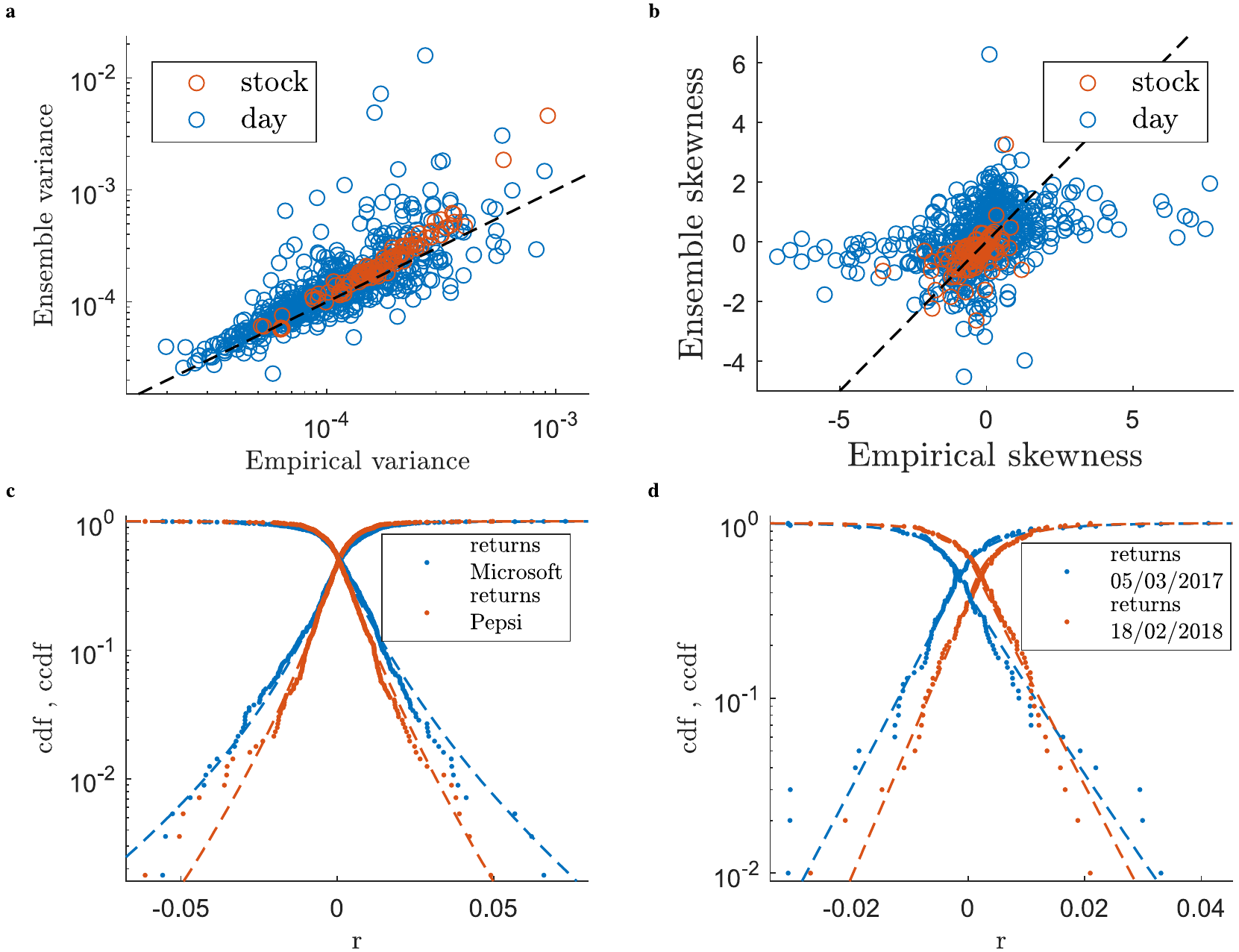}
  \caption{\textbf{Comparisons between empirical statistical properties and ensemble averages.} In these plots we demonstrate the model's ability to partially reproduce non-trivial statistical properties of the original set of time series that are not explicitly encoded as ensemble constraints. \textbf{a}) Empirical vs ensemble average values of the variances of the returns calculated for each stock (red dots) and each day (blue dots). \textbf{b}) Same plot for the skewness of the returns. \textbf{c}) Comparison between the ensemble and empirical cumulative distributions (and associated survival functions) for the returns of two randomly selected stocks (Microsoft and Pepsi Company). Dots correspond to the cumulative distribution and survival functions obtained from the empirical data. Dashed lines correspond to the equivalent functions obtained by pooling together $10^6$ time series independently generated from the ensemble. Different colours refer to different stocks as reported in the legend. Remarkably, a Kolmogorov-Smirnov test (0.01 significance) shows that 92\% of the stocks returns empirical distributions are compatible with their ensemble counterparts. \textbf{d}) Same plot for the returns of all stocks on two randomly chosen days. In this case, 82\% of daily returns empirical distributions are compatible with their ensemble counterparts (K-S test at 0.01 significance).}  
\label{fig: stock figure explained stuff}
\end{figure*}

\floatsetup[figure]{style=plain,subcapbesideposition=top}
\begin{figure*}
  \centering
\includegraphics[width=1\linewidth]{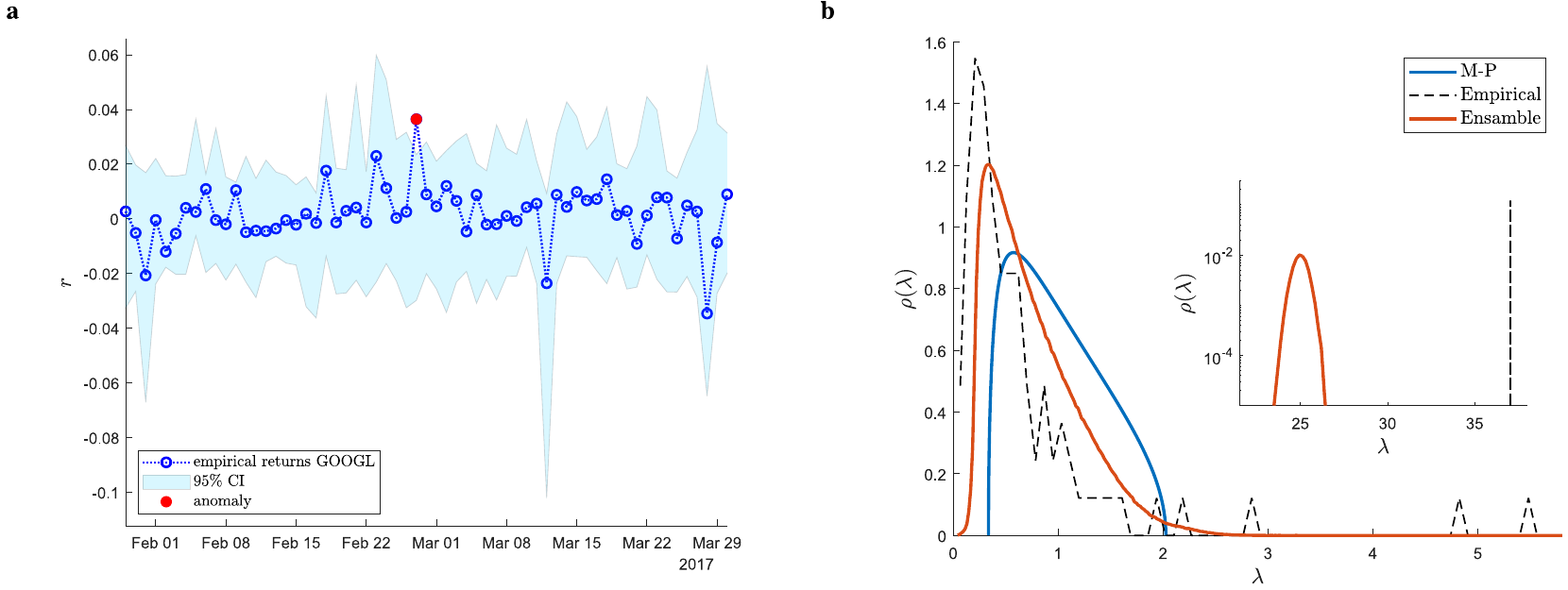}  
 \caption{\textbf{Applications of the ensemble theory we propose to a system of stocks.} \textbf{a)} Anomaly detection performed on each single trading day of a randomly selected stock (Google). A return measured on a specific day for a specific stock is marked as anomalous if it exceeds the associated 95\% confidence interval on that specific return (accounting for multiple hypothesis correction via False Coverage Rate~\cite{FCR}).  \textbf{b)} Comparison between the empirical spectrum of the estimated correlation matrix (black dashed line), its ensemble counterpart (orange line) and the one prescribed by the Marchenko-Pastur law (blue line). The inset shows the empirical largest eigenvalue (dahsed line) against the ensemble distribution for it.}
  \label{fig: applications}
\end{figure*}

\begin{table}
\vspace{0.02\textheight}
\begin{tabular}{|c|c|ccc|c|}
\hline
\multicolumn{2}{|c|}{\textbf{Returns}} & \multicolumn{3}{c|}{\textbf{Significance null hypothesis}} & \multirow{2}{*}{\textbf{\begin{tabular}[c]{@{}c@{}}median\\ rel. err.\end{tabular}}} \\ \cline{1-5}
\textbf{Stat} & \textbf{Sample} & \multicolumn{1}{c|}{\textbf{0.01-0.99}} & \multicolumn{1}{c|}{\textbf{0.05-0.95}} & \textbf{0.1-0.9} &  \\ \hline
\multirow{2}{*}{\textbf{Var}} & stock & 0.95 & 0.76 & 0.59 & 0.2 \\ \cline{2-2}
 & day & 0.88 & 0.78 & 0.69 & 0.14 \\ \cline{1-2}
\multirow{2}{*}{\textbf{Skew}} & stock & 1 & 0.98 & 0.95 & 0.13 \\ \cline{2-2}
 & day & 0.78 & 0.58 & 0.49 & 0.46 \\ \cline{1-2}
\multirow{2}{*}{\textbf{Kurt}} & stock & 0.78 & 0.61 & 0.51 & 0.60 \\ \cline{2-2}
 & day & 0.85 & 0.68 & 0.55 & 0.1 \\ \hline
\end{tabular}
\caption{Fraction of empirical moments compatible with their corresponding ensemble distribution at different significance levels specified in terms of quantiles (e.g., 0.01-0.99 denotes that the 1st and 99th percentiles of the ensemble distribution are used as bounds to determine whether the null hypothesis of an empirical moment being compatible with the ensemble distribution can be rejected or not). Note that the confidence intervals used to obtain these results have not been adjusted for multiple hypothesis testing. Doing so (e.g., via False Coverage Rate~\cite{FCR}) would further suppress the number of true positives, resulting an even larger fraction of moments being compatible with the ensemble distribution. Moments are calculated both for each stock and each trading day. In the last column, we also report, for each moment, the median relative error between the empirical value and its ensemble average.}
\label{tab: stock tab explained stuff 2}
\end{table}

\begin{table}
\begin{tabular}{|c|c|cc|}
\hline
\multirow{4}{*}{\textbf{\begin{tabular}[c]{@{}c@{}}Ratio empirical \\ aggregated pdfs\\ not rejected\\ by a K-S test\end{tabular}}} & \multirow{2}{*}{\textbf{\begin{tabular}[c]{@{}c@{}}Aggregation\\ level\end{tabular}}} & \multicolumn{2}{c|}{\textbf{\begin{tabular}[c]{@{}c@{}}K-S  test\\ significance\end{tabular}}} \\ \cline{3-4} 
 &  & \multicolumn{1}{c|}{\textbf{0.01}$\;\;$} & \textbf{0.05} \\ \cline{2-4} 
 & \textbf{stocks} & 0.92 & 0.68 \\ \cline{2-2}
 & \textbf{days} & 0.82 & 0.75 \\ \hline
\end{tabular}
\caption{Fraction of empirical return distributions (both for stocks and trading days) that are compatible with their ensemble counterparts based on Kolmogorov-Smirnov tests at different significance levels.}
\label{tab: stock tab explained stuff}
\end{table}


As an example application of the ensemble defined above, let us consider the daily returns of the  $N=100$ most capitalized NYSE stocks over $T=560$ trading days (spanning October 2016 - November 2018). In this example, the aforementioned constraints force the ensemble to preserve, on average, the number of positive and negative returns and the overall positive and negative return for each time series and for each trading day, leading to $6 (N \times T)$ constraints. When these constraints are enforced, an explicit expression for the marginal distributions can be obtained (see Appendix \ref{app:derivation}):
\begin{equation}
\label{eq: single stock prob}
    P(W_{i t} = x) = (1-P_{i t}^+) \; \lambda_{i t}^- \; e^{\lambda_{i t}^- x} \; \Theta(- x) + P_{i t}^+ \; \lambda_{i t}^+ \; e^{-\lambda_{i t}^+ x} \; \Theta(x)  \ ,
\end{equation}
where $\lambda_{i t}^\pm$ are also functions of the Lagrange multipliers (specified in Appendix \ref{app:derivation}). The above distribution allows both to efficiently sample the ensemble numerically and to obtain analytical results for several observables. Remarkably, it has been shown~\cite{masuda_mixture} that sampling from a mixture-like density such as the one in Eq.~\eqref{eq: single stock prob} can result in heavy tailed distribution, which is of crucial importance when dealing with financial data.

Figure~\ref{fig: stock figure explained stuff} and Tables~\ref{tab: stock tab explained stuff 2} and \ref{tab: stock tab explained stuff} illustrate how the above first-moment constraints translate into explanatory power of higher-order statistical properties. Indeed, in the large majority of cases, the empirical return distributions of individual stocks and trading days and their higher-order moments (variance, skewness, and kurtosis) are statistically compatible with the corresponding ensemble distributions, i.e., with the distributions of such quantities computed over large numbers ($10^6$ in all cases shown) of time series independently generated from the ensemble. Notably, this is the case without constraints explicitly aimed at enforcing such level of agreement. This, in turn, further confirms that the ensemble can indeed be exploited to perform reliable hypothesis testing by sampling random scenarios that are however closely based on the empirically available data. 

In that spirit, in Figure~\ref{fig: applications} we show an example of ex-post anomaly detection, where the original time series of a stock is plotted against the 95\% confidence intervals obtained from the ensemble for \emph{each} data point $W_{i t}$. As it can be seen, the results are non-trivial, since the returns flagged as anomalous are not necessarily the largest ones in absolute value. This is because the constraints imposed on the ensemble reflect the \emph{collective} nature of financial market movements, thus resulting in the statistical validation of events that are anomalous with respect to the overall heterogeneity present in the market.  

Following the above line of reasoning, in Figure~\ref{fig: applications} we show a comparison between the eigenvalue spectrum of the empirical correlation matrix of the data, and the average eigenvalue spectrum of the ensemble. As is well known, the correlation matrix spectrum of most complex interacting systems normally features a large bulk of small eigenvalues which is often approximated by the Marchenko-Pastur (MP) distribution \cite{marchenko_distribution} of Random Matrix Theory (i.e., the average eigenvalue spectrum of the correlation matrix of a large system of uncorrelated variables with finite second moments) \cite{plerou1999universal,laloux1999noise,livan2018introduction}, plus a few large and isolated eigenvalues that contain information about the relevant correlation structure of the system (e.g., they can be associated to clusters of strongly correlated variables \cite{livan2011fine}). As it can be seen in the Figure, the ensemble's average eigenvalue spectrum qualitatively captures the same range of the empirical spectral bulk (for reference, we also plot the MP distribution), and the ensemble distribution for the largest eigenvalue is very close to the one empirically observed, demonstrating that the main source of correlation in the market is well captured by the ensemble. Conversely, the average distance between the empirically observed largest eigenvalue and its ensemble distribution can be interpreted as the portion of the market's collective movement which cannot be explained by the constraints imposed on the ensemble.

In Appendix \ref{app:temperature}, we also apply the above ensemble approach to a dataset of weekly and hourly temperature time series recorded in North-American cities. We do this to showcase the approach's ability to capture inherent time periodicities in empirical data -- which would be very hard to capture directly -- by means of the constraints already considered in the examples above.

\section*{Applications to financial risk management}

In this section we push the examples of the previous section -- where we demonstrated the ensemble's ability to partially capture the \emph{collective} nature of fluctuations in multivariate systems -- towards real-world applications. Namely, we will illustrate a case study devoted to financial portfolio selection.

Financial portfolio selection is an optimization problem which entails allocating a fixed amount of capital across $N$ financial stocks. Typically, the goal of an investor is to allocate their capital in order to maximise the portfolio's expected return while minimising the portfolio's expected risk. When the latter is quantified in terms of portfolio variance, the solution to the optimization problem amounts to computing portfolio weights $\pi_i$ ($i = 1, \ldots, N$), where $\pi_i$ is the amount of capital to be invested in stock $i$ (note that $\pi_i$ can be negative when short selling is allowed). As is well known, these are functions of the portfolio's correlation matrix~\cite{merton1972analytic}, which reflects the intuitive notion that a well balanced portfolio should be well diversified, avoiding similar allocations of capital in stocks that are strongly correlated. The mathematical details of the problem and explicit expressions for the portfolio weights are provided in Appendix \ref{app:markowitz}.

The fundamental challenge posed by portfolio optimization is that portfolio weights have to be first computed ``in-sample'' and then retained ``out-of-sample''. In practice, this means that portfolio weights are always computed based on the correlations between stocks observed over a certain period of time, after which one observes the portfolio's realized risk based on such weights. This poses a problem, as financial correlations are known to be noisy\cite{laloux1999noise,plerou1999universal} and heavily non-stationary\cite{non-stationarity-market-livan}, so there is no guarantee that portfolio weights that are optimal in-sample will perform well in terms of out-of-sample risk.

A number of solutions have been put forward in the literature to mitigate the above problem. Most of these amount to methods to ``clean'' portfolio correlation matrices\cite{bun2016cleaning}, i.e., procedures aimed at subtracting noise and unearthing the ``true'' correlations between stocks (at least over time windows where they can be reasonably assumed to be constant). Here, we propose to exploit our ensemble approach in order to apply the same philosophy \emph{directly on financial returns} rather than on their correlation structure.

Using the same notation as in the previous section, let us assume that $\overline{W}_{it}$ represents the time-$t$ return of stock $i$ ($i = 1, \ldots, N$; $t = 1, \ldots, T$). Let us then define detrended returns $\tilde{W}_{it} = \overline{W}_{it} - \langle W_{it} \rangle$, where $\langle W_{it} \rangle$ denotes the ensemble average of the return computed from Eq.~\eqref{eq: single stock prob}. The rationale for this is to mitigate the impact of returns that may be anomalously large (in absolute value), i.e., returns whose values are markedly distant from their typical values observed in the ensemble (as in the example shown in Fig.~\ref{fig: applications}).

In Table~\ref{tab: portfolio variance} we show the results obtained by performing portfolio selection on the aforementioned detrended returns, and compare them to those obtained without detrending. Namely, we form four portfolios (two of size $N=20$ and two of size $N=50$) with the returns of randomly selected S\&P500 stocks in the period from September 2014 to October 2018, and compute their weights based on the correlations computed over non-overlapping period of lengths $T = N/q$, where $q \in (0,1)$ is the portfolio's ``rectangularity ratio'', which provides a reasonable proxy for the noisiness of the portfolio's correlation matrix (which indeed becomes singular for $q \rightarrow 1$). The numbers in the Table represent the average and $90\%$ confidence level intervals of the out-of-sample portfolio risk -- quantified in terms of variance -- computed over a number of non-overlapping time windows of $30$ days within in the aforementioned period (see caption for more details), with the first two rows corresponding to the raw returns and the two bottom rows corresponding to the detrended returns (note that in both cases out-of-sample risk is still computed on the raw returns, i.e., detrending is only performed in-sample to compute the weights). As it can be seen, detrending by locally removing the ensemble average from each return reduces out-of-sample risk dramatically, despite the examples considered here being plagued by a number of well-known potential downsides, such as small portfolio size, small time windows, and exposure to outliers (the returns used here are well fitted by power law distributions -- using the method in~\cite{clauset2009power} -- whose median tail exponent across all stocks is $\alpha = 3.9$). In Appendix \ref{app:sharpe} we report the equivalent of Table~\ref{tab: portfolio variance} for the portfolios' Sharpe ratios (i.e., the ratio between portfolio returns and portfolio variance over a time window), with qualitatively very similar results.

\begin{table}[h!]
\begin{tabular}{c|cccc|}
\cline{2-5}
 &
  \multicolumn{1}{c|}{$P_1^{20}$} &
  \multicolumn{1}{c|}{$P_2^{20}$} &
  \multicolumn{1}{c|}{$P_1^{50}$} &
  $P_2^{50}$ \\ \hline
\multicolumn{1}{|c|}{$q = 2/3$} &
  \begin{tabular}[c]{@{}c@{}}0.041\\ (0.027 , 0.091)\end{tabular} &
  \begin{tabular}[c]{@{}c@{}}0.955\\ (0.026 , 0.815)\end{tabular} &
  \begin{tabular}[c]{@{}c@{}}0.1029\\ (0.011 , 0.287)\end{tabular} &
  \begin{tabular}[c]{@{}c@{}}0.1553\\ (0.015 , 0.461)\end{tabular} \\ \cline{1-1}
\multicolumn{1}{|c|}{$q = 1/4$} &
  \begin{tabular}[c]{@{}c@{}}0.8488\\ (0.031 , 2.867)\end{tabular} &
  \begin{tabular}[c]{@{}c@{}}1.001\\ (0.022, 3.011)\end{tabular} &
  \begin{tabular}[c]{@{}c@{}}0.7846\\ (0.009 , 0.933)\end{tabular} &
  \begin{tabular}[c]{@{}c@{}}0.0938\\ (0.009 , 0.136)\end{tabular} \\ \hline
\multicolumn{1}{|c|}{$q = 2/3$} &
  \begin{tabular}[c]{@{}c@{}}0.0093\\ (0.0053 , 0.0155)\end{tabular} &
  \begin{tabular}[c]{@{}c@{}}0.0081\\ (0.0046 , 0.0124)\end{tabular} &
  \begin{tabular}[c]{@{}c@{}}0.0034\\ (0.0021 , 0.0056)\end{tabular} &
  \begin{tabular}[c]{@{}c@{}}0.0033\\ (0.0023 , 0.0053)\end{tabular} \\ \cline{1-1}
\multicolumn{1}{|c|}{$q = 1/4$} &
  \begin{tabular}[c]{@{}c@{}}0.0113\\ (0.0055 , 0.0158)\end{tabular} &
  \begin{tabular}[c]{@{}c@{}}0.0081\\ (0.0053 , 0.0111)\end{tabular} &
  \begin{tabular}[c]{@{}c@{}}0.0041\\ (0.0022 - 0.0056)\end{tabular} &
  \begin{tabular}[c]{@{}c@{}}0.0033\\ (0.0021 , 0.0054)\end{tabular} \\ \hline
\end{tabular}
\caption{Out-of-sample portfolio risk -- quantified in terms of variance -- with and without detrending the returns by subtracting their ensemble average. $P_{1,2}^N$ (with $N = 20, 50$) refer to two different portfolios made of randomly selected S\&P stocks, whereas $q = N/T$ denotes the portfolios' ``rectangularity ratio'' (i.e., the ratio between the number of stocks and the length of the in-sample time window used to compute correlations and portfolio weights). The two top rows refer to portfolios whose weights are computed based on the raw returns, whereas the two bottom rows refer to portfolios whose weights are computed based on the detrended returns. In the latter case, the detrending is only performed in-sample to compute correlations and weights, and the out-of-sample risk is computed by retaining such weights on new raw returns. The numbers reported in each case refer to the average out-of-sample risk computed over a set of 30-days long non-overlapping time windows spanning the period September 2014 - November 2018.}
\label{tab: portfolio variance}
\end{table}

In Appendix \ref{app:overfitting}, we provide details of an additional application of our ensemble approach to financial risk management, where we compute and test the out-of-sample performance of estimates of Value-at-Risk (VaR) -- the most widely used financial risk measure~\cite{jorion2000value} -- based on our ensemble approach. This application is specifically aimed at demonstrating that -- despite the large number of Lagrange multipliers necessary to calibrate the more constrained versions of our ensembles -- the approach does not suffer from overfitting issues. Quite to the contrary -- in line with the literature on configuration models for networked systems~\cite{can_ens_continuous,can_ens_nature_rev}, which are a fairly close relative of the approach proposed here -- we find the out-of-sample performance of our method to improve even when increasing the number of Lagrange multipliers quite substantially. The reason for this lies in the fact that in classic cases of overfitting one completely suppresses any in-sample variance of the model being used (e.g., when fitting $n$ points with a polynomial of order $n-1$). This is not the case, instead, with the model at hand. Indeed, being based on maximum entropy, our approach still allows for substantial in-sample variance even when building highly constrained ensembles. We illustrate this in the aforementioned example in Appendix \ref{app:overfitting} by obtaining progressively better out-of-sample VaR performance when increasing the number of constraints (and therefore or Lagrange multipliers) the ensemble is subjected to.

\section*{Relationship with Maximum Caliber principle}

Before concluding, let us point out an interesting connection between our approach and Jaynes' Maximum Caliber principle \cite{jaynes1980minimum}. It has recently been shown \cite{marzen2010equivalence} that the time-dependent probability distribution that maximizes the caliber of a two-state system evolving in discrete time can be calculated by mapping the time domain of the system as a spatial dimension of an Ising-like model. This is exactly equivalent to our mapping of a time-dependent system onto a data matrix, where the system's time dimension is mapped onto a discrete spatial dimension of the lattice representing the matrix. 

From this perspective, our ensemble approach represents a novel way to calculate and maximize the caliber of systems sampled in discrete time with a continuous number of states. This also allows to interpret some recently published results on correlation matrices in a different light. Indeed, in \cite{masuda2018configuration} the authors obtain a probability distribution on the data matrix of sampled multivariate systems starting from a Maximum Entropy ensemble on their corresponding correlation matrices. Following the steps outlined in our paper, the same results could be achieved via the Maximum Caliber principle by first mapping the time dimension of the system onto a spatial dimension of a corresponding lattice, and by then imposing the proper constraints on it.

\section*{Discussion}

In this paper we have put forward a novel formalism - grounded in the ensemble theory of statistical mechanics - to perform hypothesis testing on time series data. Whereas in physics and in the natural sciences, hypothesis testing is carried out through repeated controlled experiments, this is rarely the case in complex interacting systems, where the lack of statistical stability and controllability often hamper the reproducibility of experimental results. This, in turn, prevents from assessing whether the observations made are consistent with a given hypothesis on the dynamics of the system under study.

The framework introduced here tackles the above issues by means of a data-driven and assumptions-free paradigm, which entails the generation of ensembles of randomized counterparts of a given data sample. The only guiding principle underpinning such a paradigm is that of entropy maximization, which allows to interpret the ensemble's partition function in terms of a precise physical analogy. Indeed, as we have shown, in our framework events in a data sample correspond to fermionic particles in a physical system with multiple energy levels. In this respect, our approach markedly differs from other known methods to generate synthetic data, such as bootstrap. Notably, even though the Hamiltonians used throughout the paper correspond to non-interacting systems, and therefore the correlations in the original data are not captured in terms of interactions between particles (as is instead the case in Ising-like models), the ensemble introduced here is still capable of partially capturing properties typical of interacting systems through the `environment' the particles are embedded in, i.e., a system of coupled local temperatures and chemical potentials.

All in all, our framework is rather flexible, and can be easily adapted to the data at hand by removing or adding constraints from the ensemble's Hamiltonian. From this perspective, the number and type of constraints can be used to ``interpolate'' between very different applications. In fact, loosely constrained ensembles can serve as highly randomized counterparts of an empirical dataset of interest, and therefore can be used for statistical validation purposes, i.e., to determine which statistical properties of the empirical data can be explained away with a few basic constraints. The opposite situation is instead represented by a heavily constrained ensemble, designed to capture most of the statistical properties of the empirical data (the examples detailed in previous sections go in this direction). From a practical standpoint, this application of our approach can be particularly useful in situations where sensitive data cannot be shared between parties (e.g., due to privacy restrictions), and where sharing synthetic data whose statistical properties closely match those of the empirical data can be a very valuable alternative.

For example, the constraints in the applications showcased here (i.e., on the sums of above and below average values) result in two fermionic species of particles. More stringent constraints (e.g., on the data belonging to certain percentiles of the empirical distribution) would result in other species being added to the ensemble.

As we have shown, our framework is capable of capturing several non-trivial statistical properties of empirical data that are not necessarily associated with the constraints imposed on the ensemble. As such, it can provide valuable insight on a variety of complex systems by both allowing to test theoretical models for their structure and by allowing to uncover new information in the statistical properties that are not fully captured by the ensemble. We have illustrated some of these aspects with a financial case study, where we demonstrated that a detrending of stock returns based on our ensemble approach yields a dramatic reduction in out-of-sample portfolio risk.

Once more, let us stress that the main limitation of our approach is that of not accounting explicitly for cross-correlations or temporal correlations. As mentioned above, these could in principle be tackled by following the same analytical framework presented here, but would result in a considerably less tractable model. We aim to explicitly deal with the case of temporal correlations in future work.

\section*{Data availability}

The financial data employed in this paper are freely available to download form Yahoo Finance.
\section*{Acknowledgements}

G.L. acknowledges support from an EPSRC Early Career Fellowship in Digital Economy (Grant No. EP/N006062/1).

\section*{Author contributions statement}

Authors contribution: R.M. and G.L. designed research; R.M. performed research and analyzed data; R.M. and G.L. wrote the paper.

\section*{Additional information}

\noindent \textbf{Competing Interests:} The authors declare no conflict of interest.\\

\bibliography{sample}

\appendix

\section{Explicit calculation of the partition function}
\label{app:derivation}
We want to find a probability density function $P(W)$ on $\mathcal{W}$ such that the expectation values of a set of observables coincide with their empirical value, i.e $\langle \mathcal{O}_\ell(W) \rangle = \overline{O}_\ell$ ($\ell = 1, \ldots, L$), where $\overline{W} \in \mathcal{W}$ is the empirical set of measurements. At first, this problem may appear almost impossible to solve, given that $P(W)$ may be determined by a number of degrees of freedom way larger than the number of constraints we are imposing. However, as introduced in the main text, this can be done by using the Maximum Entropy principle, or, in other words, by adding another (functional) constraint on the probability distribution, which requires that $P(W)$ should also maximise the Gibbs entropy:
\begin{equation}\label{eq: gibbs entropy}
    S(W) = - \sum_{W \in \mathcal{W}} P(W) \, \ln  P(W) \ ,
\end{equation}
while preserving the constraints: 
\begin{equation}\label{eq: constr general}
\langle \mathcal{O}_\ell(W) \rangle = \sum_{W \in \mathcal{W}} \mathcal{O}_\ell(W) \, P(W) \; = \; \mathcal{O}_\ell(\overline{W}) = \overline{O}_\ell \ ,
\end{equation}
and the normalization:
\begin{equation}\label{eq: normal general}
    \sum_{W \in \mathcal{W}} P(W) \; = \; 1 \ .
\end{equation} 
Eqs.~\eqref{eq: gibbs entropy}-\eqref{eq: normal general} define a constrained optimization problem, whose solution is found by solving the following equation:
\begin{equation} \label{eq: opt cond}
     \frac{\partial}{\partial \, P} \left[ S + \alpha \left( 1 - \sum_{W \in \mathcal{W}} P(W) \right)+ \sum_{\ell=1}^L \beta_\ell \left( O_\ell - \sum_{W \in \mathcal{W}} \mathcal{O}_\ell(W) \, P(W) \right) \right] = 0 \ ,
\end{equation}
where, as usual in such scenarios, each constraint has been coupled with a Lagrange multiplier $\alpha,\beta_\ell$ ($\ell = 1, \ldots, L$). Defining $H(W) = \sum_\ell \beta_\ell \mathcal{O}_\ell(W)$ as the Hamiltonian of the ensemble and $Z = e^{\alpha+1} = \sum_W e^{-H(W)}$ its partition function, the solution of Eq.~\eqref{eq: opt cond} reads:
\begin{equation} \label{eq:Z}
    P(W) = \frac{e^{- H(W)}}{Z} \ .
\end{equation}
This is the general probability density function ruling the ensemble theory we are proposing. Of course, the sum $\sum_W$ on the phase space of the system used in the above equations still needs to be properly specified. 

We are going to do so while considering the Hamiltonian specified in the main paper. As pointed out above, in order to find the partition function $Z$ of the system, we just need to sum $e^{-H(W)}$ over all possible configurations, i.e., over the set of all the $N \times T$ real valued matrices $\mathcal{W}$. Recalling the notations introduced in the main text $A^{\pm} = \Theta(\pm W)$ and $w^{\pm} = \pm W \Theta (\pm W)$, we can write the sum over the phase space as follows:
\begin{equation}\label{eq: sum on config}
    \sum_{W \in \mathcal{W}}  \equiv  \prod_{i=1}^{N} \prod_{t=1}^{T} \sum_{\subalign{
       &(0,1)\\
       (A_{i t}^+,A_{i t}^-)=&(1,0)\\
       &(0,0)
    }} \; \int_{0}^{+\infty} d w^+_{i t} \; \int_{0}^{+\infty} d w^-_{i t}  \ .
\end{equation}

We can now calculate the partition function $Z$ of the ensemble:
\begin{equation}\label{eq: partition function total}
    \begin{aligned}
    Z =& \sum_{W \in \mathcal{W}} e^{-H(W)} = \\
     =& \prod_{i=1}^{N} \prod_{t=1}^{T} \sum_{\subalign{
       &(0,1)\\
       (A_{i t}^+,A_{i t}^-)=&(1,0)\\
       &(0,0)
    }} \; \int_{0}^{\infty} d w^+_{i t} \; \int_{0}^{\infty} d w^-_{i t} \;\; e^{-\left[ \left(\alpha^N_i+\alpha^T_t \right) A^+_{it} + \left( \beta^{N}_{i} + \beta^T_t \right) A^-_{it} + \left( \gamma_i^N + \gamma_t^T \right) w^+_{it} + \left( \sigma_i^{N} + \sigma_t^T \right) w^-_{it} \right]} \\
    =& \prod_{i=1}^{N} \prod_{t=1}^{T} \left( 1 + \int_{0}^{\infty} d w \; e^{- \left(\alpha^N_i+\alpha^T_t \right) - \left( \gamma_i^N + \gamma_t^T \right) w } - \int_{0}^{\infty} d w \; e^{-\left(\beta^N_i+\beta^T_t \right) + \left( \sigma_i^N + \sigma_t^T \right) w} \right)  \\
    =& \prod_{i=1}^{N} \prod_{t=1}^{T}  \left[ 1 + \frac{e^{-(\alpha^{N}_i+\alpha^{T}_t)}}{\gamma_i^{N}+\gamma_t^{T}} + \frac{e^{-(\beta^{N}_i+\beta^{T}_t)}}{\sigma_i^{N}+\sigma_t^{T}} \right] \\
    =& \prod_{i=1}^{N} \prod_{t=1}^{T} \left ( 1 + e^{\frac{\mu_{it}^1 - \epsilon_{it}}{T_{it}}} + e^{\frac{\mu_{it}^2 - \epsilon_{it}}{T_{it}}} \right ) \ ,
    \end{aligned}
\end{equation}
where all the Lagrange multipliers must be positive and we have defined the following quantities in order to make apparent the analogy with the two species fermionic gas introduced in the main text:
\begin{equation*}
    \begin{aligned}
        T_{i j} &= \frac{1}{\log{(\sigma_i^T + \sigma_j^e)} + \log{(\gamma_i^T + \gamma_j^e)}} \; , \\
        \epsilon_{i j} &= \frac{1}{2}+\frac{ T_{i j}}{2} \left( \alpha^{T}_i+\alpha^{e}_j + \beta^{T}_i+\beta^{e}_j \right) \; , \\
        \mu^2_{i j} &= \frac{k T_{i j}}{2} \left( \alpha^{T}_i+\alpha^{e}_j - \beta^{T}_i-\beta^{e}_j - \log  \frac{\sigma_i^T + \sigma_j^e}{\gamma_i^T + \gamma_j^e}  \right) = - \mu^1_{i j} \ .
    \end{aligned}
\end{equation*}

From the above partition function, via Eq.~\eqref{eq:Z} we can derive the probability density function in Eq. (5) of the main text, which quantifies the probability of drawing a specific instance $W$ from the ensemble. The quantities defining such probability distribution have a well defined physical meaning, and read as follows:
\begin{description}
    \item[$P_{it}^+ = \frac{e^{-\left(\alpha^N_i+\alpha^T_t \right)}}{\left( \gamma_i^N+\gamma_t^T \right) Z_{it}}$] Probability of observing a positive value in the $i$-th time series at time $t$
    \item[$P_{it}^- = \frac{e^{-\left (\beta^N_i+\beta^T_t \right)}}{\left( \sigma_i^N+\sigma_t^T \right) Z_{it}}$] Probability of observing a negative value in the $i$-th time series at time $t$
    \item[$1 - P_{it}^+ - P_{it}^-$] Probability of observing a missing value in the $i$-th time series at time $t$
     \item[$Q^+_{it}(w) = (\gamma_i^N+\gamma_t^T) e^{-(\gamma_i^N+\gamma_t^T) w}$] Probability distribution of a positive value $w$ for the $i$-th time series at time $t$ 
    \item[$Q^-_{it}(w) = (\sigma_i^N+\sigma_t^T) e^{-(\sigma_i^N+\sigma_t^T) w}$] Probability distribution of a negative value $w$ for the $i$-th time series at time $t$
\end{description}

When no data are missing, i.e. $(A_{i t}^+,A_{i t}^-) \neq (0,0)$, the sum defined in Eq.~\eqref{eq: sum on config} changes and, as a result, the partition function~\eqref{eq: partition function total} becomes :
\[
    Z = \prod_{i,t=1}^{N,T} Z_{i t} = \prod_{i,t=1}^{N,T}  \left[  \frac{e^{-(\alpha^{N}_i+\alpha^{T}_t)}}{\gamma_i^{N}+\gamma_t^{T}} + \frac{1}{\sigma_i^{N}+\sigma_t^{T}} \right] \;\; . 
\]
After noticing that $A_{i t}^+ = 0 \Rightarrow w_{i t}^+ = 0 \, \wedge \, w_{i t}^- > 0$, the probability of drawing from the ensemble an instance $W$ can be easily found:
\begin{equation}\label{eq: final prob stock}
\begin{aligned}
 P&(W) =  \prod_{i,t=1}^{N,T} \left[ P_{it}^+ \;\; Q_{it}^+(w^+_{it})  \right]^{A^+_{it}} \left[ P_{it}^- \;\; Q_{it}^-(w^-_{it}) \right]^{1-A^+_{it}} \ ,
\end{aligned}
\end{equation}
where the quantities in the above expression are defined as those above.

Looking at Eq.~\eqref{eq: final prob stock}, we can understand how we have obtained Eq. (6) in the main text. In order to simulate a drawing of a set of time series $W$ from the ensemble, we first need to construct a ``topology'' of positive events by placing a positive event in entry $W_{i t}$ with probability $P_{i t}^+$ and a negative event otherwise. Then we need to place a weight $W_{i t}=x$ using one of the two exponential distributions $Q_{it}^\pm$ defined above, depending on the type of event that was assigned to $W_{i t}$. This procedure is encompassed by the hyperexponential distribution in Eq. (6) of the main text, which can be obtained via the standard generating function approach, and whose parameters read $\lambda_{it}^+ = (\gamma_i^N + \gamma_i^T)^{-1}$, and $\lambda_{it}^- = (\sigma_i^N + \sigma_i^T)^{-1}$.


\section{Application to a set of temperature time series} 
\label{app:temperature}

\floatsetup[figure]{style=plain,subcapbesideposition=top}
\begin{figure*}[]
  \centering
  \sidesubfloat[]{\includegraphics[width=0.32\linewidth]{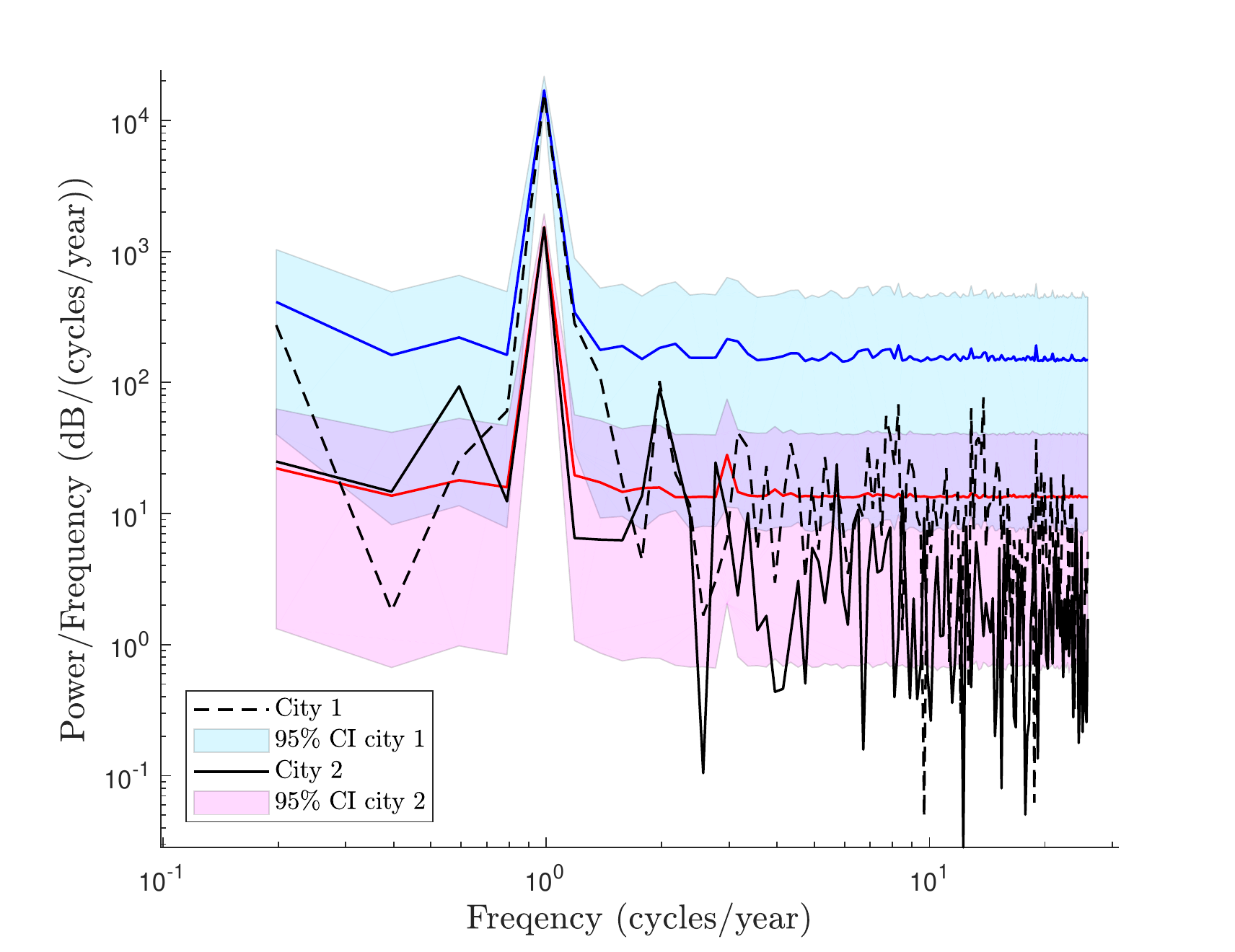}}
  \sidesubfloat[]{\includegraphics[width=0.32\linewidth]{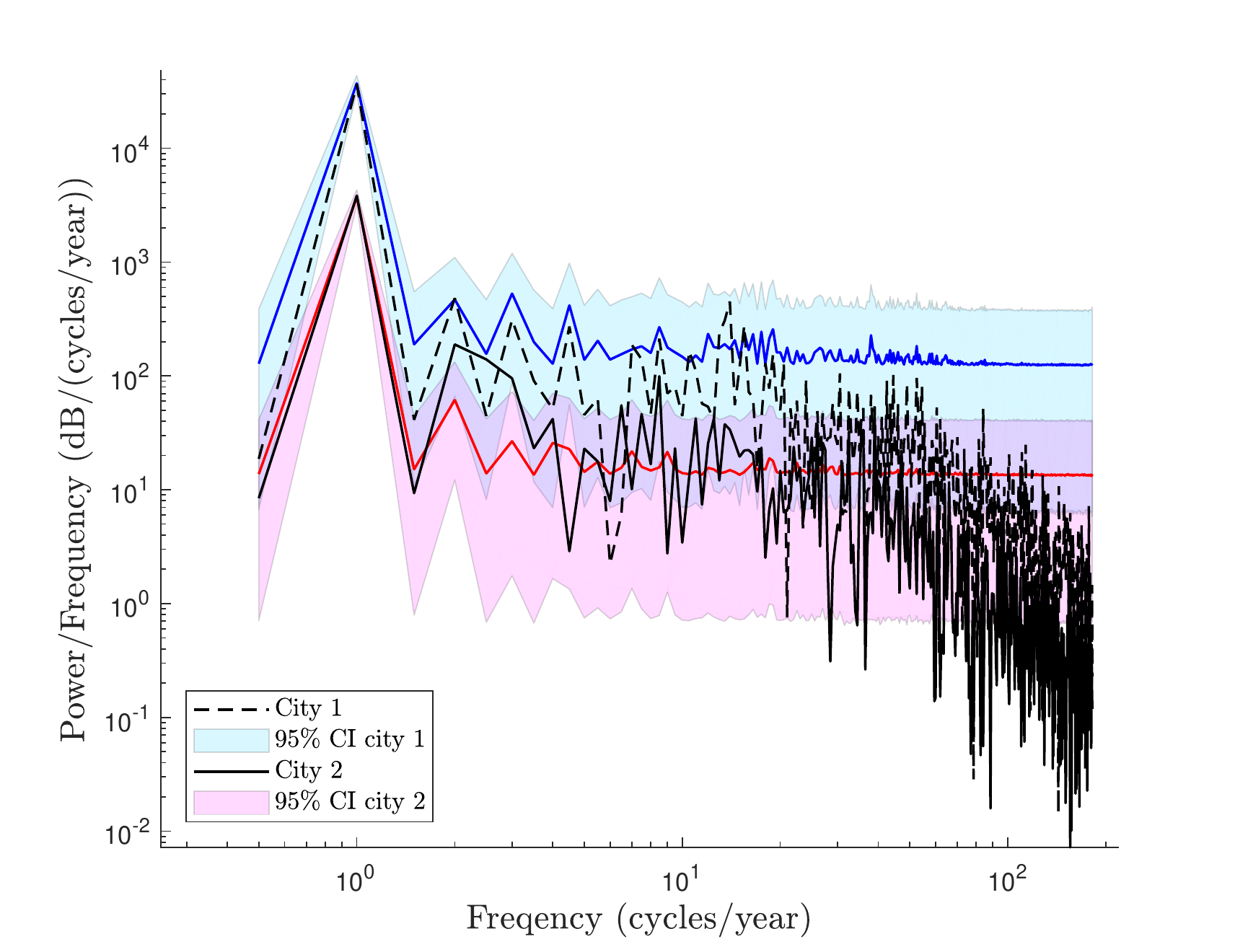}}
  \sidesubfloat[]{\includegraphics[width=0.32\linewidth]{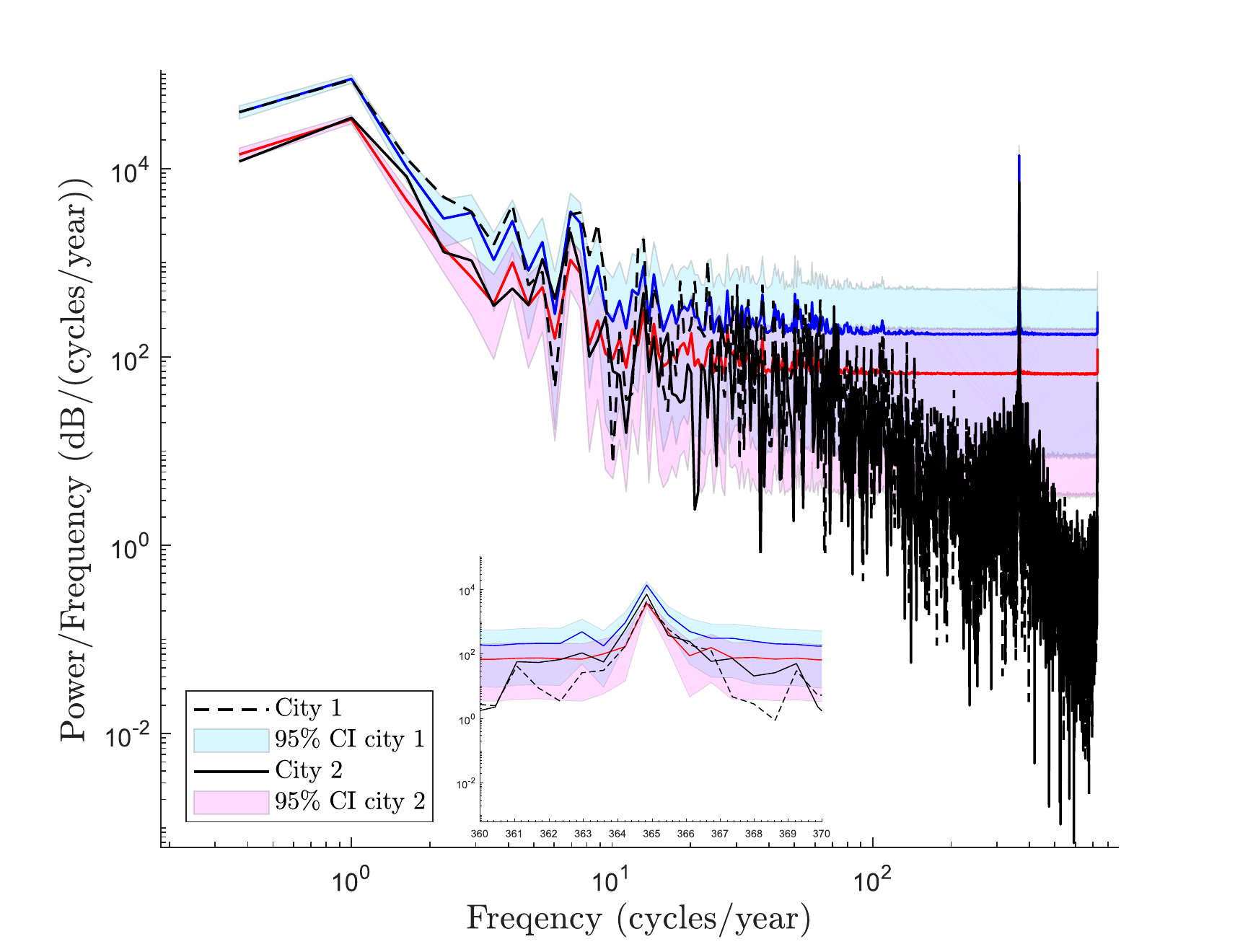}}
  \caption{\textbf{Ability of the esemble to preserve periodicities in the data.} \textbf{a)} Empirical power spectrum of weekly temperatures against the average ensemble spectrum for two different cities (city 1 is Boston and city 2 is Los Angeles). \textbf{b)} Same plot for daily temperatures. \textbf{c)} Same plot for 8 hours temperatures.  }
  \label{fig: periodicity temperature}
\end{figure*}

We now apply the framework introduced in the main text to sets of time series featuring temperatures recorded at different frequencies (week/day/8 hours) in $ N = 30$ different North American cities \footnote{Vancouver, Portland, San Francisco, Seattle, Los Angeles, San Diego, Las Vegas, Phoenix, Albuquerque, Denver, San Antonio, Dallas, Houston, Kansas City, Minneapolis, Saint Louis, Chicago, Nashville, Indianapolis, Atlanta, Detroit, Jacksonville, Charlotte, Miami, Pittsburgh, Toronto, Philadelphia, New York, Montreal, Boston} (weekly data range from July 2013 to July 2018, daily data range from July 2016 to July 2018, 8 hour data range from January 2017 to July 2018). We do so in order to test the ability of our ensemble approach to capture the main features of time series whose most relevant statistical properties are markedly different from those of financial returns, which we instead studied in the main paper. In particular, our main focus will be on the ability of the ensemble to capture the periodicities that characterize temperature data at different time scales. 

As done in the main text, we indicate as $\overline{W}$ the $N \times T$ data matrix (with $T = 264, 730, 2321$ in the case of temperatures recorded at the weekly, daily, and 8 hour frequency, respectively) with values rescaled to have zero mean, and we indicate as $W$ any generic instance drawn from the corresponding ensemble. We also redefine here for convenience the matrices $A^{\pm} = \Theta(\pm W)\, , \, w^{\pm} = \pm W \Theta (\pm W)$. The ensemble we are going to use is fully specified by the $6(N+T)$ constraints enforced in the following Hamiltonian (there are no missing data, which leads to $2 (N+T)$ fewer constraints with respect to the general formulation outlined in the main paper):
\begin{equation}\label{eq: hamiltonian app}
   H(W) =  \sum_{i=1}^N \sum_{t=1}^T \left [ \left(\alpha^N_i+\alpha^T_t \right) A^+_{it} + \left( \gamma_i^N + \gamma_t^T \right) w^+_{it} + \left( \sigma_i^{N} + \sigma_t^T \right) w^-_{it}  \right] \ ,
\end{equation}
leading to the partition function:
\begin{equation}\label{eq: partition function}
    Z = \prod_{i=1}^N \prod_{t=1}^T Z_{i t} = \prod_{i=1}^N \prod_{t=1}^T  \left[  \frac{e^{-(\alpha^{N}_i+\alpha^{T}_t)}}{\gamma_i^{N}+\gamma_t^{T}} + \frac{1}{\sigma_i^{N}+\sigma_t^{T}} \right] \;\; . 
\end{equation}

In Figure~\ref{fig: periodicity temperature} we show that, independently from the frequency at which temperatures are sampled, the average ensemble power spectral density captures well the relevant frequencies that characterize the empirical time series of each city. Indeed, as can be seen from panels \textbf{a} and \textbf{b}, the ensemble power spectra based on the data recorded at the weekly and daily frequency perfectly capture the six-months periodicity associated with the seasons' cycle. Panel \textbf{c} shows that the same frequency is also captured in the data recorded every 8 hours, and that when calibrating the ensemble on such data, the power spectrum also perfectly captures the daily frequency associated with the day-night cycle (see inset).

\begin{figure*}[]
  \centering
  \sidesubfloat[]{\includegraphics[width=0.42\linewidth]{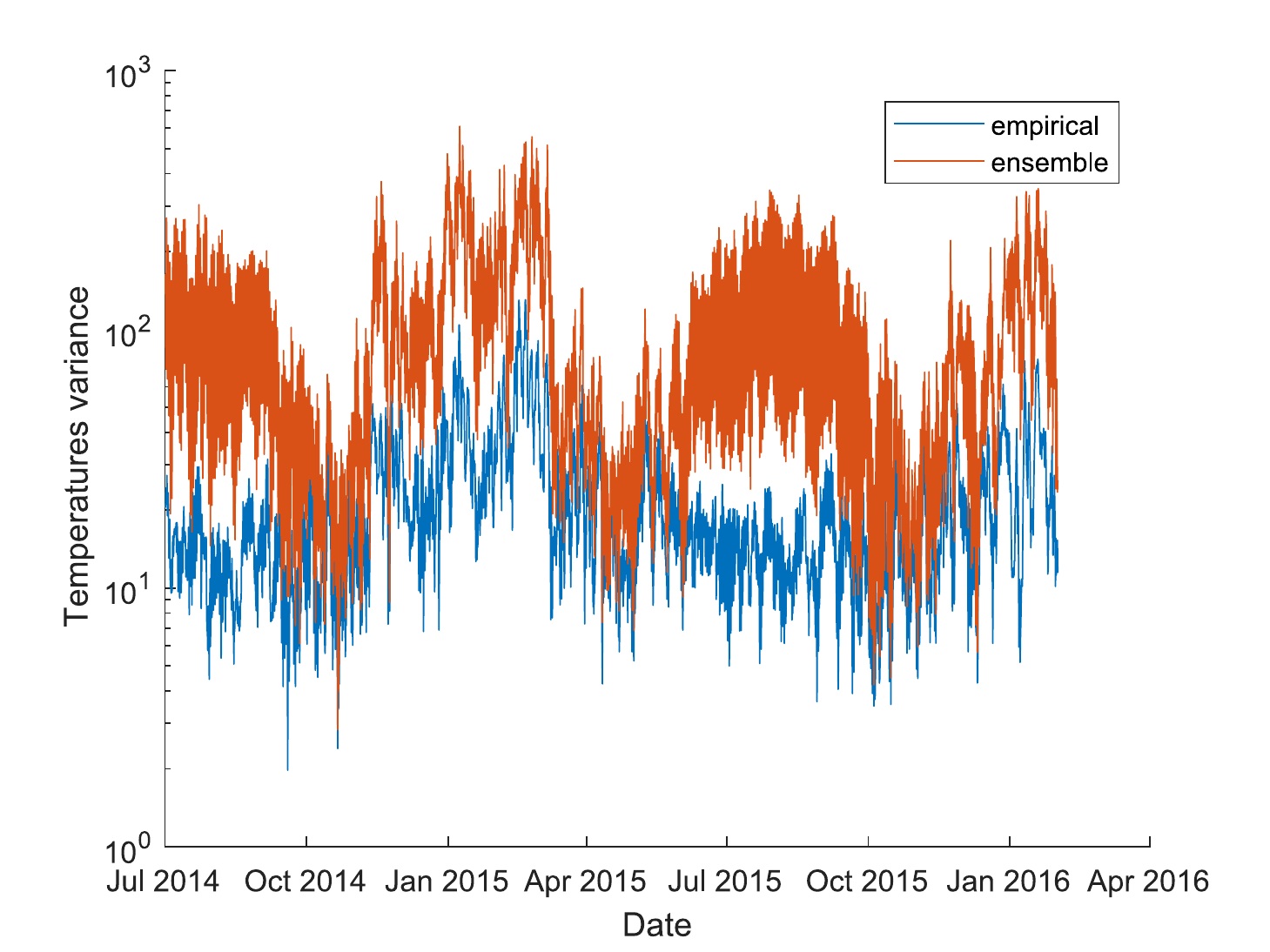}}
  \sidesubfloat[]{\includegraphics[width=0.42\linewidth]{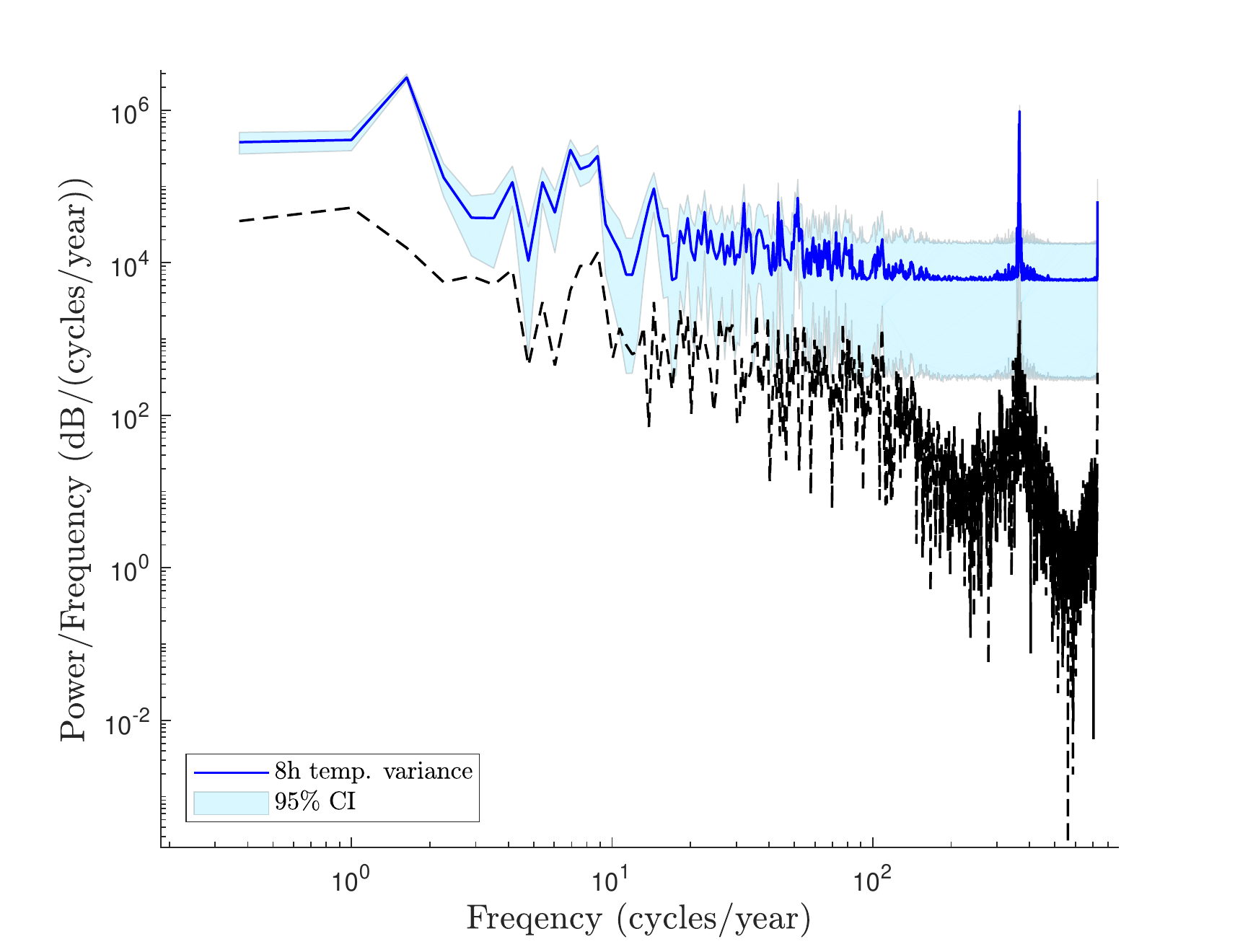}} \\  
  \sidesubfloat[]{\includegraphics[width=0.42\linewidth]{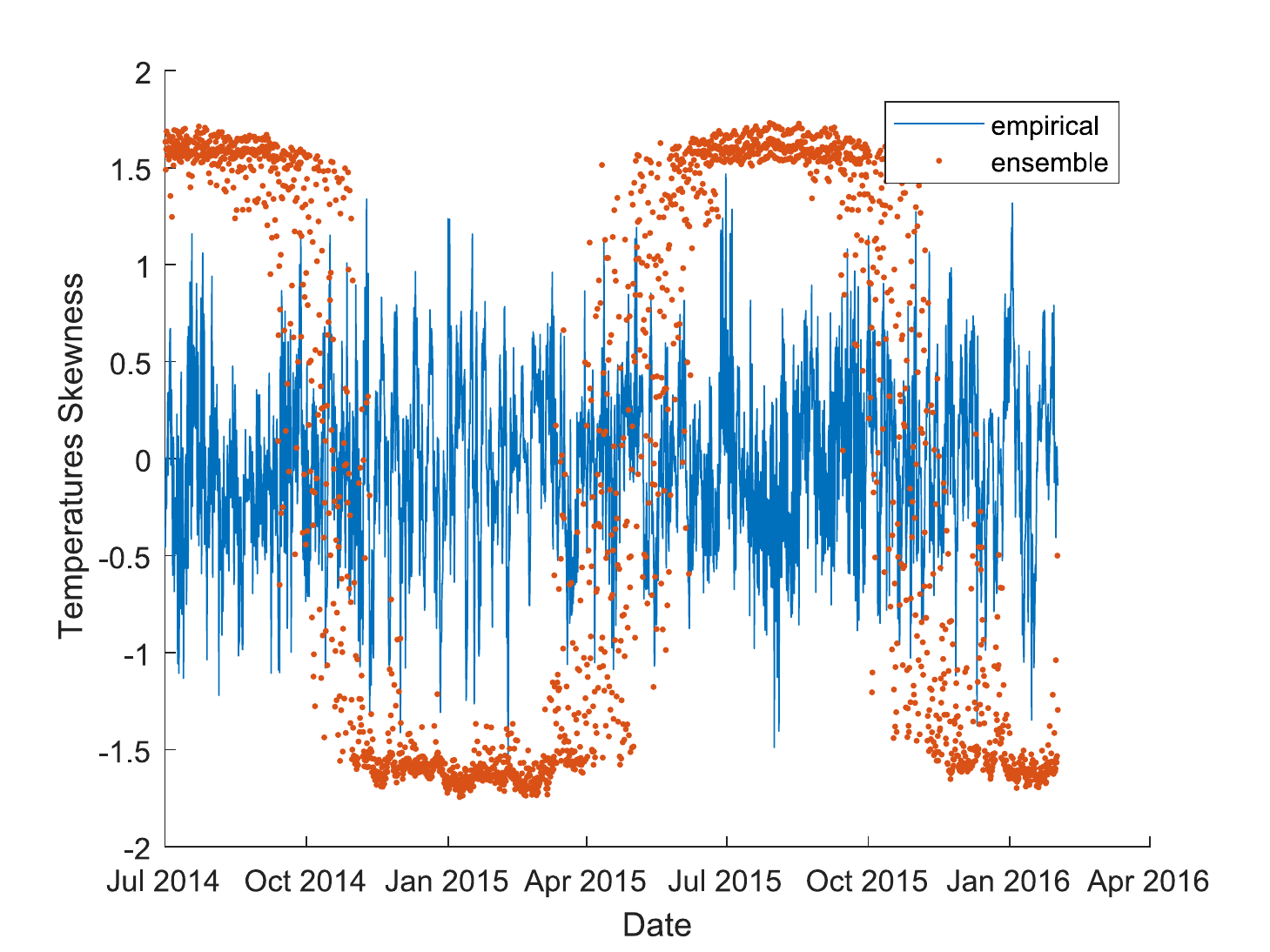}}
  \sidesubfloat[]{\includegraphics[width=0.42\linewidth]{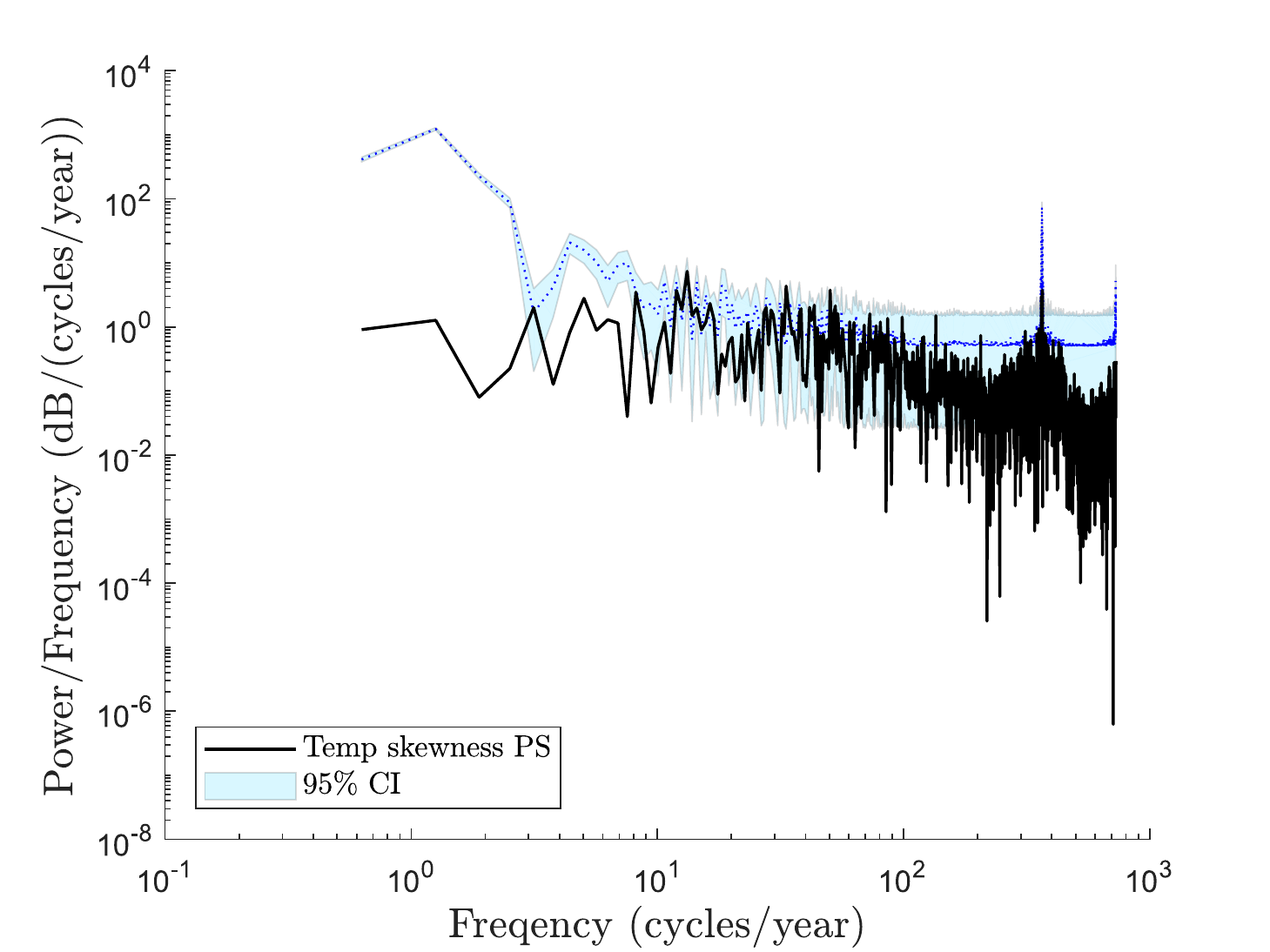}} \\
  \caption{\textbf{Ability of the ensemble to preserve periodicities in the data.} \textbf{a)} Variance of the temperatures recorded at 8 hour intervals across all 30 cities (the blue line denotes empirical values, the orange one denotes the ensemble average). \textbf{b)} Comparison between the empirical spectrum of the 8-hours temperature variance across cities (dashed line) and the ensemble spectrum (blue line). \textbf{c)} Skewness of the temperatures recorded at 8 hour intervals across all 30 cities (the blue line denotes empirical values, the orange one denotes the ensemble average). \textbf{d)} Comparison between the empirical spectrum of the 8-hours temperature skewness across cities (dashed line) and the ensemble spectrum (blue line).}
  \label{fig: periodicity moments temperature}
\end{figure*}

In Fig.~\ref{fig: periodicity moments temperature} we expand the above analysis to the periodicities of moments. Panel \textbf{a} shows the empirical daily variance of temperatures recorded across the 30 cities mentioned above against the corresponding ensemble average. At first sight, the latter seems to be largely uncorrelated from the former. Yet, the corresponding power spectrum shown in panel \textbf{b} highlights that the relevant frequencies in the data (six months and one day) are captured very well, although the ensemble places additional power on such frequencies. 

\floatsetup[figure]{style=plain,subcapbesideposition=top}

A somewhat similar phenomenon is shown in panels \textbf{c} and \textbf{d}, which show the daily skewness computed across all cities and its corresponding power spectra. Once again, the average ensemble spectrum places more power on the six-months and daily frequencies with respect to the empirical one. This results in a clearly discernible oscillating pattern, which significantly deviates from the empirical behavior. Nevertheless, these results are interesting. Indeed, as it can be seen in panel \textbf{c} positive (negative) skewness values take place during the summer (winter) months, as a reflection of higher (lower) average temperatures. Although this is a fairly trivial example, it highlights how the ensemble approach can reveal stylized trends that are genuinely informative about the dynamics of the system under study.


\section{Optimal portfolio selection} 
\label{app:markowitz}

Let us consider a matrix $r_{it}$ ($i = 1,\ldots,N$, $t = 1,\ldots,T$) of daily financial returns, and let $C$ be the corresponding correlation matrix (i.e., $C_{ij}$ denotes the Pearson correlation coefficient between stocks $i$ and $j$, computed over the time period $[1,T]$). The optimal portfolio problem then amounts to solving the following optimization problem for a vector $\boldsymbol{\pi} = (\pi_1, \ldots, \pi_N) \in \mathbb{R}^N$:
\begin{equation} \label{eq:portfolio_variance}
\underset{\boldsymbol{\pi}}{\min} \sum_{i,j=1}^N C_{ij} \pi_i \pi_j
\end{equation}
subject to
\begin{equation} \label{eq:portfolio_constraints}
\sum_{i=1}^N \pi_i \mu_i = \mu \ ; \qquad \sum_{i=1}^N \pi_i = 1 \ ,
\end{equation} 
where Eq.~\eqref{eq:portfolio_variance} expresses the minimization of the portfolio variance, while the equations in \eqref{eq:portfolio_constraints} express constraints on the expected returns ($\mu_i$ denotes the expected out-of-sample return of stock $i$, while $\mu$ denotes the portfolio's desired expected return) and on the available capital (conventionally set to one unit). The expected returns $\mu_i$ are computed -- as is often customary -- based on mean reversion, i.e., assuming that the return on day $t+1$ will be minus the return on day $t$.

With the above positions, the solution to the optimization problem reads (see Ref. [37] in the main paper)
\begin{equation}
\pi_i(\mu) = \sum_{j=1}^N C^{-1}_{ij} (\ell(\mu) + \mu g(\mu)) \ ,
\end{equation}
where
\begin{eqnarray}
&& \ell(\mu) = \frac{c - b\mu}{a c - b^2} \ , \qquad g(\mu) = \frac{a\mu - b}{a c - b^2} \\
&& a = \sum_{i,j=1}^N C_{ij}^{-1} \ , \qquad b = \sum_{i,j=1}^N C_{ij}^{-1} \mu_j \ , \qquad c = \sum_{i,j=1}^N C_{ij}^{-1} \mu_i \mu_j \ .
\end{eqnarray}


\section{Sharpe ratios of portfolios in Table 3} 
\label{app:sharpe}

In the Table below we report the Sharpe ratios (defined as the ration between the portfolio return and portfolio variance over a time interval) for the same portfolios considered in Table 1 of the main paper.

\begin{table}[h!]
\begin{tabular}{c|cccc|}
\cline{2-5}
 &
  \multicolumn{1}{c|}{$P_1^{20}$} &
  \multicolumn{1}{c|}{$P_2^{20}$} &
  \multicolumn{1}{c|}{$P_1^{50}$} &
  $P_2^{50}$ \\ \hline
\multicolumn{1}{|c|}{$q = 2/3$} &
  \begin{tabular}[c]{@{}c@{}}0.032\\ (-0.19 , 0.31)\end{tabular} &
  \begin{tabular}[c]{@{}c@{}}0.009\\ (-0.22 , 0.23)\end{tabular} &
  \begin{tabular}[c]{@{}c@{}}-0.077\\ (-0.25 , 0.15)\end{tabular} &
  \begin{tabular}[c]{@{}c@{}}-0.022\\ (-0.22 , 0.28)\end{tabular} \\ \cline{1-1}
\multicolumn{1}{|c|}{$q = 1/4$} &
  \begin{tabular}[c]{@{}c@{}}-0.013\\ (-0.21 , 0.12)\end{tabular} &
  \begin{tabular}[c]{@{}c@{}}-0.022\\ (-0.26, 0.23)\end{tabular} &
  \begin{tabular}[c]{@{}c@{}}0.011\\ (-0.19 , 0.20)\end{tabular} &
  \begin{tabular}[c]{@{}c@{}}-0.037\\ (-0.27 , 0.15)\end{tabular} \\ \hline
\multicolumn{1}{|c|}{$q = 2/3$} &
  \begin{tabular}[c]{@{}c@{}}0.042\\ (-0.14 , 0.22)\end{tabular} &
  \begin{tabular}[c]{@{}c@{}}0.041\\ (-0.16 , 0.22)\end{tabular} &
  \begin{tabular}[c]{@{}c@{}}0.081\\ (-0.17 , 0.34)\end{tabular} &
  \begin{tabular}[c]{@{}c@{}}0.054\\ (-0.17, 0.25)\end{tabular} \\ \cline{1-1}
\multicolumn{1}{|c|}{$q = 1/4$} &
  \begin{tabular}[c]{@{}c@{}}0.035\\ (-0.23 , 0.26)\end{tabular} &
  \begin{tabular}[c]{@{}c@{}}0.035\\ (-0.17 , 0.24)\end{tabular} &
  \begin{tabular}[c]{@{}c@{}}0.073\\ (-0.16 , 0.32)\end{tabular} &
  \begin{tabular}[c]{@{}c@{}}0.062\\ (-0.13 , 0.34)\end{tabular} \\ \hline
\end{tabular}
\end{table}

As it can be seen, applying the ``cleaning'' procedure outlined  in the main paper has beneficial effects also in this case, leading to increased average Sharpe ratios and reduced uncertainty around them across board.


\section{Testing for overfitting with an application on the estimation of Value-at-Risk} 
\label{app:overfitting}

In this section we expand on the financial application of our ensemble approach discussed in the main paper, with a specific focus on potential overfitting issues. Namely, the number of Lagrange multipliers the ensemble depends on increase linearly with the number of constraints one wants to enforce. For instance, the multivariate case detailed in the main paper (Eq.~(7)) depends on $8(N+T)$, which, for small numbers of variables $N$ and small sample sizes $T$, can be of the same order of magnitude of the number of data points ($N \times T$) used to calibrate the ensemble. This, in turn, may raise concerns about potential overfitting issues.

We tackle the above issue by showcasing the approach's performance when using it to compute out-of-sample Value-at-Risk (VaR) estimates based on increasingly constrained (and therefore parametrized) versions of the ensemble. VaR is the most widely adopted measure of financial risk. For a given significance level $\alpha$ it simply refers to the $1-\alpha$ quantile of the return distribution of a certain financial stock or portfolio. The simplest procedure to estimate VaR is via historical estimation, which amounts to computing the in-sample $1-\alpha$ quantile of a financial time series of interest. However, due to non-stationarities, historical estimates are known to typically be unreliable out-of-sample, and there is a vast literature devoted to enhancing historical estimates with Monte Carlo simulations and other techniques to generate synthetic scenarios. In the following, we will apply a similar line of reasoning by computing VaR estimates as quantiles of the distributions generated by our ensemble starting from an empirical time series of interest, and we will assess their out-of-sample performance based on a number of standard statistical tests.

In the following, we consider two financial time series of length $T = 1000$ and $T = 1500$ days corresponding, respectively, to BNP returns from May 7, 2008 to March 13, 2013, and to S\&P Index returns from May 7, 2008 to July 26, 2014. For each time series we proceed to compute VaR estimates with a rolling window approach. Namely, we compute an in-sample VaR estimate over a time window $[t_0,t_0+\tau]$, with $\tau = 150$ days, and assess its out of sample performance on day $t_0+\tau+1$. We do this based on the following three versions of our ensemble approach:

\begin{itemize}
\item[{\bf Model $M1$}] -- This corresponds to a loosely constrained ensemble based on the single time series case discussed in the paper, where we only constrain the ensemble to preserve the empirical time series' variance and the cumulative values of the data falling within each pair of adjacent quartiles (denoted as $\overline{M}_{\xi_i}$ in the main paper, with $\xi_i = 0.25, 0.5, 0.75$). Overall, these correspond to 4 constraints and Lagrange multipliers.

\item[{\bf Model $M2$}] -- This corresponds to a deliberately highly parametrized model based on an adaptation of the multiple time series case. Namely, let us consider the 150 returns of interest to compute a new risk estimate and let us denote them as $r_1, \ldots, r_{150}$. We then form a $25 \times 126$ (which roughly amount to the length of a trading month and half of a trading year, respectively) matrix with such returns with the following circulant structure

\begin{equation*}
R = 
\begin{pmatrix}
r_{25} & r_{26} & \cdots & r_{150} & \epsilon \\
r_{24} & r_{25} & \cdots & r_{149} &  r_{150} \\
\vdots  & \vdots  & \ddots & \vdots & \vdots  \\
r_1 & r_2 & \cdots & r_{125} &  r_{126}
\end{pmatrix} \ .
\end{equation*}
 
The quantity $\epsilon$ in the upper-right entry of the matrix denotes the unknown out-of-sample return on day 151. We assume as possible values for it $\epsilon = \pm \min |r_t|$, and then generate the corresponding ensemble constraining it to preserve the cumulative positive and negative values for each row and column (denoted respectively as $\overline{S}_i^\pm$ and $\overline{R}_t^\pm$ in the main paper), which correspond to $2(25+126) = 302$ constraints and Lagrange multipliers\footnote{It can be shown that as long as the matrix $R$'s sizes $L_1$ and $L_2$ are not multiple of each other, then such constraints are all linearly independent. In the case of linear dependence, the effective number of constraints decreases by at most $\max(L_1,L_2)$, which still amounts to an over-parametrized model.}.  We generate the ensembles for both aforementioned values of $\epsilon$ and combine the two resulting distributions for them in order to compute a VaR estimate for the return on day 151.
 
\item[{\bf Model $M3$}] -- The same as model $M2$ with additional constraints on the number of positive and negative returns recorded in each column (the equivalent of the quantity denoted as $\overline{M}_t^\pm$ in the main paper). In addition to the constraints mentioned above, this gives a total of $3(25+126) = 453$ constraints and Lagrange multipliers.
 
\end{itemize}

Models $M2$ and $M3$ are highly constrained (and therefore highly parametrized) ones, as they force the corresponding ensemble to preserve a very large number of local properties of the time series.

\begin{table}[h!]
\begin{tabular}{|c|c|c|c|}
\hline
\multirow{2}{*}{\textbf{$\alpha$}} & \multicolumn{3}{c|}{\textbf{Tests passed}} \\ \cline{2-4} 
                & $M_3$ & $M_2$ & $M_1$ \\ \hline
90\%    & 6     & 4     & 4     \\ \hline
95\%  & 8     & 6     & 5     \\ \hline
99\%   & 8     & 6     & 6     \\ \hline
99.99\% & 8     & 8     & 8     \\ \hline
\end{tabular}
\caption{Number of tests passed by out-of-sample VaR estimates for the BNP stock at different significance levels $\alpha$.}
\label{tab:bnp}
\end{table}


\begin{table}[h!]
\begin{tabular}{|c|c|c|c|}
\hline
\multirow{2}{*}{\textbf{$\alpha$}} & \multicolumn{3}{c|}{\textbf{Tests passed}} \\ \cline{2-4} 
                & $M_3$ & $M_2$ & $M_1$ \\ \hline
90\%    & 5     & 4     & 4     \\ \hline
95\%  & 8     & 7     & 6     \\ \hline
99\%     & 8     & 7     & 7     \\ \hline
99.99\% & 8     & 8     & 7     \\ \hline
\end{tabular}
\caption{Number of tests passed by out-of-sample VaR estimates for the S\&P Index at different significance levels $\alpha$.}
\label{tab:sp}
\end{table}

We calibrate the above models on all time windows of length $\tau = 150$ days starting on days $t_0 = 1, 2, \ldots, T-151$ and compute out-of-sample VaR estimates for each of them, resulting in $849$ estimates for BNP and $1349$ estimates for the S\&P Index, respectively. We then pool such estimates for both time series and assess their out-of-sample performance by means of 8 standard tests widely adopted in the financial literature. These are the traffic light, binomial, proportion of failures, time until first failure, conditional coverage, conditional coverage independence, time between failures, and time between failures independence tests (see, e.g., Nieppola, O., \emph{Backtesting Value-at-Risk Models} for their definitions).

The results -- reported as the number of tests passed -- are shown in Table~\ref{tab:bnp} for BNP and in Table~\ref{tab:sp} for the S\&P Index, for varying significance levels $\alpha$. As it can be seen, the out-of-sample performance systematically improves when increasing the number of constraints, regardless of the significance level, even when pushing these to numbers close to the number of available data points. Remarkably, all tests are passed when using model $M3$ at significance $95\%$ or higher.

\end{document}